\documentstyle{amsart}

\numberwithin{equation}{section}

\newtheorem{theorem}{Theorem}[section]

\newtheorem{lemma}[theorem]{Lemma}

\newtheorem{proposition}[theorem]{Proposition}

\theoremstyle{remark}
\newtheorem{remark}[theorem]{Remark}   
          
\newtheorem{remarknon}{Remark}

\newcommand{\diag}{\operatorname{diag}}
\newcommand{\Ad}{\operatorname{Ad}}
\newcommand{\tr}{\operatorname{tr}}

\begin{document}

\title[Monopoles and the Gibbons-Manton metric]
      {Monopoles and the Gibbons-Manton metric}

\author{Roger Bielawski}
\subjclass{53C25, 81T13}
\address{ Max-Planck-Institut f\"ur Mathematik\\ Gottfried-Claren-Strasse 26\\
53225 Bonn, Germany}

\email{rogerb@@mpim-bonn.mpg.de}

\begin{abstract} 
We show that, in the region where monopoles are well separated, the $L^2$-metric on the moduli space of  $n$-monopoles is exponentially close to the $T^n$-invariant hyperk\"ahler metric proposed by Gibbons and Manton. The proof is based on a description of the Gibbons-Manton metric as a metric on a certain moduli space of solutions to Nahm's equations, and on twistor methods. In particular, we show how the twistor description of monopole metrics determines the asymptotic metric. 
   \end{abstract}

\maketitle

The moduli space $M_n$ of (framed) static $SU(2)$-monopoles of charge $n$, i.e. solutions to Bogomolny equations $d_A\Phi=\ast F$, carries a natural hyperk\"ahler metric \cite{AtHi}. The geodesic motion in this metric is a good approximation to the dynamics of low energy monopoles \cite{Man,Stu}. For the charge $n=2$ the metric has been determined  explicitly by Atiyah and Hitchin \cite{AtHi}, and it follows from their explicit formula that when the two monopoles are well separated, the metric becomes (exponentially fast) the Euclidean Taub-NUT metric with a negative mass parameter. It was also shown by N. Manton \cite{Man1} that this asymptotic metric can be determined by treating well-separated monopoles as dyons. The equations of motion for a pair of dyons in ${\Bbb R}^3$ are found to be equivalent to the equations for geodesic motion on Taub-NUT space.
\par
For an arbitrary charge $n$, it was shown in \cite{BielCrelle} that, when the individual monopoles are well-separated, the $L^2$-metric is close (as inverse of the separation distance) to the flat Euclidean metric. Gibbons and Manton \cite{GM} have then calculated the Lagrangian for the motion of $n$ dyons in ${\Bbb R}^3$ and shown that it is equivalent to the Lagrangian for geodesic motion in a hyperk\"ahler metric on a torus bundle over the configuration space $\tilde{C}_n({\Bbb R}^3)$. This metric is $T^n$-invariant and has a simple algebraic form. Gibbons and  Manton have conjectured, by analogy with the $n=2$ case, that the exact $n$-monopole metric differs from their metric by an exponentially small amount as the separation gets large. We shall prove this conjecture here. 
\par
Our strategy is as follows. We construct certain moduli space $\tilde{M}_n$ of solutions to Nahm's equations which carries a $T^n$-invariant hyperk\"ahler metric. Using twistor methods we identify this metric as the Gibbons-Manton metric. Finally, we show that the metrics on $\tilde{M}_n$ and $M_n$ are exponentially close. This proof adapts equally well to the asymptotic behaviour of $SU(N)$-monopole metrics with maximal symmetry breaking, as will be shown elsewhere.
\par
The asymptotic picture can be explained in the twistor setting. We recall that a monopole is determined (up to framing) by a curve $S$ - the spectral curve - in $T{\Bbb C}P^1$, which satisfies certain conditions \cite{Hit}. One of these is triviality of the line bundle $L^{-2}$ over $S$, and a nonzero section of this bundle is the other ingredient needed to determine the metric \cite{Hurt,AtHi}.
Asymptotically we have now the following situation. When the individual monopoles become well separated the spectral curve of the $n$-monopole degenerates (exponentially fast) into the union of spectral curves $S_i$ of individual monopoles, while the section of $L^{-2}$ becomes (also exponentially fast) $n$ meromorphic sections of $L^{-2}$ over the individual $S_i$. The zeros and poles of these sections occur only at the intersection points of the curves $S_i$. This information (and the topology of the asymptotic region of $M_n$) is, as we show in the last section, sufficient to conclude that the asymptotic metric is the Gibbons-Manton metric.
\medskip

The article is organized as follows. In sections 1 and 2 we recall the definitions of the Gibbons-Manton and monopole metrics. In section 3 we introduce the moduli space $\tilde{M}_n$ of solutions to Nahm's equations and give heuristic arguments why the metric on  $\tilde{M}_n$ should be exponentially close to the monopole metric. In section 4, as a preliminary step to study $\tilde{M}_n$ we introduce yet another moduli space of solutions to Nahm's equations, somewhat simpler than $\tilde{M}_n$. In that section we also discuss the relation with Kronheimer's metrics on $G^{\Bbb C}/T^{\Bbb C}$, where $G$ is a compact semisimple Lie group and $T\leq G$ is a maximal torus. In section 5 we identify $\tilde{M}_n$ as a differential, complex, and finally complex-symplectic manifold. In section 6 we calculate the twistor space of $\tilde{M}_n$ and identify its hyperk\"ahler metric as the Gibbons-Manton metric. In section 7 we finally show that the monopole metric and the metric on $\tilde{M}_n$ are exponentially close. Short section 8 shows how one can read off the Gibbons-Manton metric, as the asymptotic form of the monopole metric, from the twistor description of the latter.

\section{The Gibbons-Manton metric\label{zero}}

The Gibbons-Manton metric \cite{GM} is an example of $4n$-dimensional (pseudo)-hyperk\"ahler metric admiting a tri-Hamiltonian (hence isometric) action of the $n$-dimensional torus $T^n$.  Such metrics have particularly nice properties and were studied by several authors \cite{LR,HKLR,PP}. The Gibbons-Manton metric was described  as a hyperk\"ahler quotient of a flat quaternionic vector space by Gibbons and Rychenkova in \cite{GR}. We recall here this description, which we slightly modify to better suit our purposes. 
We start with flat hyperk\"ahler metrics $g_1$ and $g_2$ on $M_1=\bigl(S^1\times {\Bbb R}^3\bigr)^n$ and $M_2={\Bbb H}^{n(n-1)/2}$. We consider a pseudo-hyperk\"ahler metric on the product manifold $M=M_1\times M_2$ given by
$ g=g_1-g_2$. The complex structures on ${\Bbb H}$ are given by the right multiplication by quaternions $i,j,k$. The metric $g_1$ is invariant under the obvious action (by translations) of $T^n=(S^1)^n$ and the metric $g_2$ is invariant under the left diagonal action of $T^{n(n-1)/2}$. We consider a homomorphism $\phi:T^{n(n-1)/2}\rightarrow T^n$ given by
$$(t_{ij})_{i<j}\mapsto \left( \prod_{j=i+1}^{n}t_{ij}\prod_{j=1}^{i-1} t_{ji}^{-1}\right)_{i=1,\ldots,n}.$$
This defines an action of  $T^{n(n-1)/2}$ on $M=M_1\times M_2$ by $t\cdot(m_1,m_2)=(\phi(t)\cdot m_1, t\cdot m_2)$. Gibbons and Rychenkova have shown that the hyperk\"ahler quotient of $(M,g)$ by this action of $T^{n(n-1)/2}$ is the Gibbons-Manton metric.

We remark that, if we choose coordinates $(t_i,{\bf x}_i)$ on $M_1$, $t_i\in S^1$ and ${\bf x}_i\in {\Bbb R}^3$, and quaternionic coordinates $q_{ij}$, $i<j$, on ${\Bbb H}^{n(n-1)/2}$, then the moment map equation are:
\begin{equation} \frac{1}{2}q_{ij}i\bar{q}_{ij}={\bf x}_i-{\bf x}_j.\label{momGM} \end{equation}
As long as ${\bf x}_i\neq {\bf x}_j$ for $i\neq j$, the torus $T^{n(n-1)/2}$ acts freely on the zero-set of the moment map. The quotient of this set by $T^{n(n-1)/2}$ is a smooth hyperk\"ahler manifold which we denote by $M_{\scriptscriptstyle GM}$. The action of $T^n$ on $M_1$ induces a free tri-Hamiltonian action on $M_{\scriptscriptstyle GM}$ for which the moment map is just $({\bf x}_1,\ldots,{\bf x}_n)$.
This makes $M_{\scriptscriptstyle GM}$ into a $T^n$-bundle over the configuration space $\tilde{C}_n({\Bbb R}^3)$ of $n$ distinct points in ${\Bbb R}^3$. We shall now determine this bundle. We recall that a basis of $H_2\bigl(\tilde{C}_n({\Bbb R}^3),{\Bbb Z}\bigr)$ is given by the $n(n-1)/2$ $2$-spheres
\begin{equation}S_{ij}^2=\{({\bf x}_1,\dots,{\bf x}_n)\in{\Bbb R}^3\otimes {\Bbb R}^n; |{\bf x}_i-{\bf x}_j|={\it const},\enskip {\bf x}_k={\it const}\enskip\text{if}\enskip k\neq i,j\}\label{basis}\end{equation}
 where $i<j$. We have
\begin{proposition} The hyperk\"ahler moment map for the action of $T^n$ makes   $M_{\scriptscriptstyle GM}$ into a $T^n$-bundle over $\tilde{C}_n({\Bbb R}^3)$ determined by the element $(s_1,\ldots,s_n)$ of \linebreak $H^2\bigl(\tilde{C}_n({\Bbb R}^3),{\Bbb Z}^n\bigr)$ given by $$s_k(S_{ij}^2)=\begin{cases} -1 &\text{if $k=i$}\\1 &\text{if $k=j$}\\0 &\text{otherwise.}\end{cases}$$\label{bundle0}\end{proposition}
\begin{pf} From the formula \eqref{momGM} it follows that restricting the 
bundle to a fixed $S_{ij}^2$ is equivalent to considering the case $n=2$. 
In other words $s_k(S_{ij}^2)=0$ if $k\neq i,j$ and we have to consider only one quaternionic coordinate $q_{ij}$. The zero-set of the moment map is $\frac{1}{2}q_{ij}i\bar{q}_{ij}={\bf x}_i-{\bf x}_j$ and the circle $S^1$ by which we quotient acts by $t\cdot\bigl(q_{ij},(t_i,{\bf x}_i),(t_j,{\bf x}_j) \bigr)=\bigl(tq_{ij},(tt_i,{\bf x}_i),(t^{-1}t_j,{\bf x}_j) \bigr)$. The quotient can be obtained by setting $t_i=1$ and the induced action of the $i$-th generator $s_i$ of $T^n$ is then given by left multiplication by $s_i^{-1}$ on $q_{ij}$. Since the map $q_{ij}\rightarrow \frac{1}{2}q_{ij}i\bar{q}_{ij}$ with the left action of $S^1$ on $\{q_{ij}\in {\Bbb H}; |q_{ij}|=1\}$ is the Hopf bundle, it follows that $s_i(S^2_{ij})=-1$. A similar argument shows that $s_j(S^2_{ij})=1$. 
\end{pf}
 
In particular, $(\overline{t},\overline{\bf x})=(t_i,{\bf x}_i)$ form local coordinates on $M_{\scriptscriptstyle GM}$. The metric tensor can be then written in the form \cite{PP}:
$$g=\Phi d\overline{\bf x}\cdot d\overline{\bf x}+\Phi^{-1}(d\overline{t}+A)^2,$$
where the matrix $\Phi$ and the $1$-form $A$ depend only on the ${\bf x}_i$ and satisfy certain linear PDE's. In particular, $\Phi$ determines the metric. For the Gibbons-Manton metric 
$$4\Phi_{ij}=\begin{cases}1-\sum_{k\neq i}\frac{1}{\|{\bf x}_i-{\bf x}_k\|} & \text{if $i=j$}\\ \frac{1}{\|{\bf x}_i-{\bf x}_j\|} & \text{if $i\neq j$}.\end{cases}$$

\section{Nahm's equations and monopole metrics \label{one}}

We shall recall in this section the description of the $L^2$-metric on the moduli space of charge $n$ $SU(2)$-monopoles in terms of Nahm's equations. A proof that the Nahm transform \cite{Nahm,Hit} between the two moduli spaces is an isometry was given by Nakajima in \cite{Nak}. 

\par  
 One starts with the space ${\cal A}$ of quadruples $(T_0,T_1,T_2,T_3)$ of smooth ${\frak u}(n)$-valued functions on $(-1,1)$ such that $T_1,T_2,T_3$ have simple poles at $\pm 1$ with residues $\frac{1}{2}\rho(\sigma_i)$, $i=1,2,3$, where $\rho:{\frak su}(2)\rightarrow {\frak u}(n)$ is the standard irreducible $n$-dimensional representation of ${\frak su}(2)$ and $\sigma_i$ are the Pauli matrices.
 Equipped with the $L^2$-norm (given by a biinvariant inner product on ${\frak u}$(n)),  ${\cal A}$ becomes a flat quaternionic affine space. There is an  isometric and triholomorphic action of the gauge group ${\cal G}$ of $U(n)$-valued functions $g:[-1,1]\rightarrow U(n)$ which are $1$ at $\pm 1$:

\begin{eqnarray} T_0&\mapsto & \Ad(g)T_0-\dot{g}g^{-1}\nonumber\\ T_i&\mapsto & \Ad(g)T_i\;,\;\;\qquad i=1,2,3.\label{action}\end{eqnarray}

 The zero-set
of the hyperk\"ahler moment map for this action is then described by {\em Nahm's equations} \cite{Nahm}:
\begin{equation}\dot{T}_i+[T_0,T_i]+\frac{1}{2}\sum_{j,k=1,2,3}\epsilon_{ijk}[T_j,T_k]=0\;,\;\;\;\;i=1,2,3.\label{Nahm}\end{equation}
The quotient of the space of solutions by ${\cal G}$ is the a smooth hyperk\"ahler manifold $M_n$ of dimension $4n$. By the above mentioned result of Nakajima, $M_n$ is the moduli space of (framed) charge $n$ $SU(2)$-monopoles. With respect to any complex structure $M_n$ is biholomorphic to the space of based rational maps of degree $n$ on  ${\Bbb C}P^1$ \cite{Don}.\newline
If we replace $U(n)$ by $=SU(n)$ (resp. by $PSU(n)$) in the above description, we obtain the moduli space of strongly centered (resp. centered) $SU(2)$-monopoles of charge $n$.

\begin{remark} A similar construction can be done for any compact Lie group $G$. We require $\rho:{\frak su}(2)\rightarrow {\frak g}$ to be a Lie algebra homomorphism whose image lies in the regular part of ${\frak g}$. We obtain  a smooth hyperk\"ahler manifold of dimension $4\,{\rm rank}\,G$ which can be identified with a totally geodesic submanifold of certain moduli space of $SU(N)$-monopoles (with a minimal symmetry breaking). Alternatively, as a complex manifold, it is a desingularization of $\left({\frak h}^{\Bbb C}\times T^{\Bbb C}\right)/W$ where $T^{\Bbb C}$ is a maximal torus in $G^{\Bbb C}$, ${\frak h}^{\Bbb C}$ its Lie algebra, and $W$ the corresponding Weyl group \cite{BielJLMS}.  
\end{remark}

The tangent space to $M_n$ can be described as the space of solutions to the linearized Nahm's equations and satisfying the condition of being orthogonal (in the $L^2$-metric) to vectors arising from infinitesimal gauge transformations. In other words the tangent space to $M_n$ at a solution $(T_0,T_1,T_2,T_3)$ can be identified  with the set of solutions $(t_0,t_1,t_2,t_3)$ to the following system of linear equations:
\begin{equation}\begin{array}{c} \dot{t}_0+[T_0,t_0]+[T_1,t_1]+[T_2,t_2]+[T_3,t_3]=0,\\ 
\dot{t}_1+[T_0,t_1]-[T_1,t_0]+[T_2,t_3]-[T_3,t_2]=0,\\
\dot{t}_2+[T_0,t_2]-[T_1,t_3]-[T_2,t_0]+[T_3,t_1]=0,\\  
\dot{t}_3+[T_0,t_3]+[T_1,t_2]-[T_2,t_1]-[T_3,t_0]=0.\end{array}\label{tangent}\end{equation}
The metric is defined by
\begin{equation}\|(t_0,t_1,t_2,t_3)\|^2=\frac{1}{2}\int_{-1}^1\sum_0^3\|t_i\|^2 \label{metric}.\end{equation}
The three anti-commuting complex structures can be seen by writing a tangent vector as $t_0+it_1+jt_2+kt_3$.\bigskip

\section{The asymptotic moduli space\label{two}}

We shall now construct a one-parameter family of moduli spaces $\tilde{M}_n(c)$, $c\in {\Bbb R}$, of solutions to Nahm's equations carrying (pseudo-)hyperk\"ahler metrics.
We shall see later on that  these metrics are the Gibbons-Manton metric with different mass parameters. 
\par
 We consider the subspace $\Omega_1$ of exponentially fast decaying functions in $C^1[0,\infty]$, i.e.:
\begin{equation}\Omega_1=\left\{f:[0,\infty]\rightarrow {\frak u}(n); \exists_{\eta >0}\sup_t\left(e^{\eta t}\|f(t)\|+e^{\eta t}\|df/dt\|\right)<+\infty\right\}.\label{omega}\end{equation}
As in the previous section, $\rho:{\frak su}(2)\rightarrow {\frak u}(n)$ is the standard irreducible $n$-dimensional representation of ${\frak su}(2)$ (in particular, $\rho(\sigma_1)$ is a diagonal matrix). We denote by ${\frak h}$ the (Cartan) subalgebra of ${\frak u}(n)$ consisting of diagonal matrices.\newline
Let $\tilde{{\cal A}}_n$ be the space of $C^1$-functions $(T_0,T_1,T_2,T_3)$  defined on $(0,+\infty]$ and satisfying (cf. \cite{Kron}):
\begin{itemize}
\item[(i)] $T_1,T_2,T_3$ have simple poles at $0$ with $\mbox{res}\,T_i=\frac{1}{2}\rho(\sigma_i)$;
\item[(ii)] $T_i(+\infty)\in {\frak h}$  for $i=0,\ldots,3$;
\item[(iii)] $(T_1(+\infty),T_2(+\infty),T_3(+\infty))$ is a regular triple, i.e. its centralizer is ${\frak h}$;
\item[(iv)]  $\left(T_i(t)-T_i(+\infty)\right)\in \Omega_1$ for $i=0,1,2,3$.\end{itemize}

Next we shall define the relevant gauge group. 
The Lie algebra of our gauge group ${\cal G}(c)$ is the space of $C^2$-paths $\rho:[0,+\infty)\rightarrow {\frak u}(n)$  such that
\begin{itemize}
\item[(i)] $\rho(0)=0$ and $\dot{\rho}$ has a limit in ${\frak h}$ at $+\infty$;
\item[(ii)] $(\dot{\rho}-\dot{\rho}(+\infty))\in \Omega_1$, and $[\tau,\rho]\in\Omega_1$ for any regular element $\tau\in{\frak h}$;
\item[(iii)] $c\dot{\rho}(+\infty)+\lim_{t\rightarrow +\infty} (\rho(t)-t\dot{\rho}(+\infty))=0$.
\end{itemize}
It is the Lie algebra of the Lie group 
$$\begin{aligned} {\cal G}(c)=\left\{g:[0,+\infty)\rightarrow\right. & U(n); \,g(0)=1, \;s(g):=\lim\dot{g}g^{-1}\in {\frak h},\;(\tau-\Ad(g)\tau)\in \Omega_1,\\ & (\dot{g}g^{-1}-s(g))\in \Omega_1,\;\left.\exp(cs(g))\lim\left(g(t)\exp(-ts(g))\right)=1\right\}.\end{aligned}$$

\begin{remarknon} The last condition in the definition of ${\cal G}(c)$ means that $g(t)$ is asymptotic to $\exp(ht-ch)$ for some diagonal $h$.\end{remarknon} 
We introduce a family of metrics on $\tilde{{\cal A}}_n$. Let $(t_0,t_1,t_2,t_3)$ be a tangent vector to the space $\tilde{{\cal A}}_n$ at a point $(T_0,T_1,T_2,T_3)$. The functions $t_i$ are now regular at $0$, $i=0,\ldots,3$. We put   
\begin{equation}\|(t_0,t_1,t_2,t_3)\|_c^2=c\sum_0^3\|t_i(+\infty)\|^2+\int_0^{+\infty}\sum_0^3\left(\|t_i(s)\|^2-\|t_i(+\infty)\|^2\right)ds. \label{smetric}\end{equation}

We observe that the group ${\cal G}(c)$ acting by (\ref{action}) preserves the metric $\|@,\cdot@,\|_c$ and the three complex structure of the flat hyperk\"ahler manifold $\tilde{{\cal A}}_n$. We define $\tilde{M}_n(c)$ as the (formal) hyperk\"ahler quotient of $\tilde{{\cal A}}_n$ by ${\cal G}(c)$ (with respect to the metric $\|@,\cdot@,\|_c$). The zero set of the moment map is given by the equations (\ref{Nahm}) (here  the condition (iii) in the definition of $\mbox{Lie}({\cal G}(c))$ is essential) and so $\tilde{M}_n(c)$ is defined as the moduli space of solutions to Nahm's equations:
$$\tilde{M}_n(c)=\left\{\text{solutions to (\ref{Nahm}) in}\enskip \tilde{{\cal A}}_n\right\}/{\cal G}(c).$$

\begin{remarknon}  If $c>0$, then the metric (\ref{smetric}) on $\tilde{M}_n(c)$ will be seen to be positive definite if  $(T_1(+\infty),T_2(+\infty)$, $T_3(+\infty))$ is sufficiently far from the walls of Weyl chambers. On the other hand, if $c<0$, then the metric will be shown to be everywhere negative definite. Therefore, for $c<0$ we should really replace $\|@,\cdot@,\|_c$ with its negative; it is, however more convenient to consider the metrics $\|@,\cdot@,\|_c$. 
\par
 We observe that sending a solution $T_i$ to the solution $rT_i(rt)$ for any $r>0$ induces a homothety of factor $r$ between $\tilde{M}_n(c)$ and $\tilde{M}_n(rc)$.\newline
\end{remarknon}

Before we begin the detailed study of $\tilde{M}_n(c)$, let us explain why we expect this metric to be exponentially close to the monopole metric. It is known \cite{BielAGAG} that the solutions to Nahm's equations on $(0,2)$ corresponding to a well-separated monopole are exponentially close to being constant away from the boundary points (i.e. on any $[\epsilon,2-\epsilon]$). The same is true for solutions on the half line $(0,+\infty)$: as long as the triple $(T_1(+\infty), T_2(+\infty), T_3(+\infty))$ is regular, the solutions are exponentially close to being constant away from $0$ \cite{Kron} (it is helpful to notice that the space of regular triples is the same as the space $\tilde{C}_n\bigl({\Bbb R}^3\bigr)$ of distinct points in ${\Bbb R}^3$). Our strategy is to take two solutions, on half-lines $(0,\infty)$ and $(-\infty,2)$ with the same values at $\pm \infty$, cut them off at $t=1$ and use this non-smooth solution on $(0,2)$ (with correct boundary behaviour) to obtain an exact solution to the monopole Nahm data. The exact solution will differ from the approximate one by an exponentially small amount. Furthermore the part of the half-line solutions which we have cut off is exponentially close to being constant and, for $c=1$, contributes an exponentially small amount to the metric $\|\,\cdot\,\|_c$ (all estimates are uniform and can be differentiated). This can be seen from the fact that we can rewrite \eqref{smetric} as 
\begin{equation}\|(t_0,t_1,t_2,t_3)\|_c^2=\int_0^{c}\sum_0^3\|t_i(s)\|^2+\int_c^{+\infty}\sum_0^3\left(\|t_i(s)\|^2-\|t_i(+\infty)\|^2\right)ds.\label{newmetric}\end{equation}
The first term, together with the corresponding term for the solution on $(-\infty,2)$, is exponentially close to the monopole metric (for $c=1$).

\section{Moduli space of regular semisimple adjoint orbits \label{modorbits}}

In order to obtain information about $\tilde{M}_n(c)$ we need to consider first  another moduli space of solutions to Nahm's equations, defined analogously, except that we require the solutions to be smooth at $t=0$. This space, which can be defined for an arbitrary compact Lie group $G$, is of some interest as all hyperk\"ahler structures on $G^{\Bbb C}/T^{\Bbb C}$ (here  $T^{\Bbb C}$ is a maximal torus) due to Kronheimer \cite{Kron} can be obtained from it as hyperk\"ahler quotients (see Theorem \ref{orbits} below). A reader who is primarily interested in monopoles should think of $G$ as $U(n)$.
\par
Let us  first recall how Kronheimer constructs hyperk\"ahler metrics on $G^{\Bbb C}/T^{\Bbb C}$. Let ${\frak h}$ be the Lie algebra of $T^{\Bbb C}$ and let $(\tau_1,\tau_2,\tau_3)\in {\frak h}^3$ be a regular triple, i.e. one whose centralizer is ${\frak h}$. For a fixed $\eta>0$, consider the Banach space
$$\Omega_1^\eta=\left\{f:[0,\infty]\rightarrow {\frak g}; \sup_t\left(e^{\eta t}\|f(t)\|+e^{\eta t}\|df/dt\|\right)<+\infty\right\}$$
with the norm $\|f\|= \sup_t\left(e^{\eta t}\|f(t)\|+e^{\eta t}\|df/dt\|\right)$. Define ${\cal A}^{\eta}(\tau_1,\tau_2,\tau_3)$ as the space  of $C^1$-functions $(T_0,T_1,T_2,T_3):(0,+\infty]\rightarrow {\frak g}$ which satisfy:
$$ \{T_0(t),\left(T_i(t)-\tau_i\right);i=1,2,3\}\subset \Omega_1^\eta.$$
Define also
 ${\cal G}^\eta$ by replacing $\Omega_1$ with $\Omega_1^\eta$ in the definition of ${\cal G}$ given in the previous section. Kronheimer shows then that for small enough $\eta$
$$M(\tau_1,\tau_2,\tau_3)=\left\{\text{solutions to (\ref{Nahm}) in}\enskip {\cal A}^{\eta}(\tau_1,\tau_2,\tau_3)\right\}/{\cal G}^\eta$$
equipped with the $L^2$ metric is a smooth hyperk\"ahler manifold, diffeomorphic to $G^{\Bbb C}/T^{\Bbb C}$. Futhermore, if $(\tau_2,\tau_3)$ is regular, then $M(\tau_1,\tau_2,\tau_3)$ is biholomorphic, with respect to the complex structure $I$, to the complex adjoint orbit of $\tau_2+i\tau_3$.
\par
We observe that the union  of all $M(\tau_1,\tau_2,\tau_3)$ has a natural topology and it is, in fact, a smooth manifold. We shall show now that there is a $T$-bundle over this union which carries a (pseudo)-hyperk\"ahler metric.
We define the space ${\cal A}_G$ by omitting the condition (i) in the definition of $\tilde{\cal A}_n$ in the previous section. Instead we require that the $T_i$ are smooth at $t=0$ for $i=0,1,2,3$. We define $M_G(c)$, $c\in {\Bbb R}$, as the (formal) hyperk\"ahler quotient of ${\cal A}_G$ by ${\cal G}(c)$ with respect to the metric (\ref{smetric}). We have:    
\begin{proposition} $M_G(c)$ equipped with the metric (\ref{smetric}) is a smooth hyperk\"ahler manifold. The tangent space at a solution $(T_0,T_1,T_2,T_3)$ is described by the equations (\ref{tangent}).\label{smooth} \end{proposition}
We remark that the metric \ref{smetric} may be degenerate at some points. However the hypercomplex structure is defined everywhere.
\begin{pf}
  Define $M_G^\eta(c)$ by replacing $\Omega$ with $\Omega^\eta$ in the definition of $M_G(c)$. By the exponential decay property of solutions to Nahm's equations (\cite{Kron}, Lemma 3.4), a neighbourhood of a particular element in $M_G(c)$ is canonically identified with its neighbourhood in $M_G^\eta(c)$ for small enough $\eta$. Therefore we can use the transversality arguments of \cite{Kron}, Lemma 3.8 and Proposition 3.9 (with a slight modification due to condition (iii) in the definition of ${\rm Lie}({\cal G}(c))$) to deduce the smoothness. The fact that the metric is hyperk\"ahler is, formally, the consequence of the fact that $M_G(c)$ is a hyperk\"ahler quotient. One can, in fact, check directly that the three K\"ahler forms are closed. We shall also, later on, identify the complex structures and the complex symplectic forms proving their closedness.\end{pf} 

We observe now that the action on ${\cal A}_G$ of gauge transformations which are asymptotic to $\exp(-th+\lambda h)$, $h\in {\frak h}$, $\lambda\in {\Bbb R}$, induce a free isometric action of $T=\exp({\frak h})$ on $M_G(c)$.
In fact this action is tri-Hamiltonian and a simple calculation shows
\begin{proposition} The hyperk\"ahler moment map $\mu=(\mu_1,\mu_2,\mu_3)$ for the action of $T$ on $M_G(c)$ is given by
$\mu_i(T_0,T_1,T_2,T_3)=T_i(+\infty)$ for $i=1,2,3$.  \qed\label{moment}\end{proposition}
As an immediate corollary we have: 
\begin{theorem} Let $(\tau_1,\tau_2,\tau_3)$ be a regular triple in ${\frak h}^3$.
The hyperk\"ahler quotient $\mu^{-1}(\tau_1,\tau_2,\tau_3)/T$ of $M_G(c)$ by the torus $T$ is isometric to Kronheimer's $M(\tau_1,\tau_2,\tau_3)$. \qed\label{orbits}\end{theorem}
We have also a tri-Hamiltonian action of $G$ on $M_G(c)$ given by the gauge transformations with arbitrary values at $t=0$. The hyperk\"ahler moment map for this action is $(T_1(0),T_2(0),T_3(0))$.  
\par
We have two other group actions on $M_G(c)$. There is a free isometric and triholomorphic action of the Weyl group $W=N(T)/T$ given by the gauge transformations which become constant (and in $W$) exponentially fast.\newline
Finally there is a free isometric $SU(2)$-action which rotates the complex structures. As a consequence it has a globally defined K\"ahler potential for each K\"ahler form  (cf. \cite{HKLR}). The potential for $\omega_2$ (or $\omega_3$) is given by the moment map for the action of a circle in $SU(2)$ which preserves $I$. This is easily seen to be
$$K_J=c\sum_{i=2}^3\|T_i(+\infty)\|^2+\int_0^{+\infty}\sum_{i=2}^3 \Bigl(\|T_i(s)\|^2-\|T_i(+\infty)\|^2\Bigr)ds.$$

\begin{remark} There is a similar (pseudo)-hyperk\"ahler manifold with a torus action such that the hyperk\"ahler quotients by this torus are isometric to Kronheimer's ALE-metrics on the minimal resolution of a given Kleinian singularity ${\Bbb C}^2/\Gamma$ \cite{KronALE}. This manifold is defined as $M_G$ except that the $T_i$ have poles at $t=0$ with the residues defined by a subregular homomorphism ${\frak su}(2)\rightarrow {\frak g}$ (cf. \cite{BielJLMS,BielAGAG2}).\end{remark}

\begin{remark} One can observe that $M_G(0)$ is a cone metric (with the ${\Bbb R}_{>0}$-action given by $T_i(t)\mapsto rT_i(rt)$) and in fact, it is an ${\Bbb H}^\ast$-bundle over a pseudo-quaternion-K\"ahler manifold (cf. \cite{Swann}).\end{remark}

\section{$\tilde{M}_n(c)$ as a manifold\label{twoandhalf}}

We now return to the space $\tilde{M}_n(c)$ defined in section \ref{two}. Our first task is to show that this space is smooth. We shall show that $\tilde{M}_n(c)$ is a smooth hyperk\"ahler quotient of the product of the space $M_{U(n)}(c-1)$ considered in the previous section and of another moduli space of solutions to Nahm's equations. This latter space, denoted by $N_n$, is given by ${\frak u}(n)$-valued solutions to Nahm's equations defined on $(0,1]$ smooth at $t=1$ and with the same poles as $\tilde{M}_n(c)$ at $t=0$. The gauge group consists of gauge transformations which are identity at $t=0,1$. Equipped with the metric (\ref{metric}) this is a smooth hyperk\"ahler manifold \cite{BielJLMS,Dan1}. It admits a tri-Hamiltonian action of $U(n)$ given by gauge transformations with arbitrary values at $t=1$. In addition, we consider the space $M_{U(n)}(c-1)$ defined in the previous section. We identify it this time with the space of solutions on $[1,+\infty]$ via the map $T_i(t)\mapsto T_i(t+1)$ (so that the gauge transformations behave now, near $+\infty$, as elements of ${\cal G}(c)$).

It is easy to observe that the space $\tilde{M}_n(c)$ is the hyperk\"ahler quotient of $N_n\times M_{U(n)}(c-1)$ by the diagonal action of $U(n)$ (cf. \cite{BielJLMS}; the moment map equations simply match the functions $T_1,T_2,T_3$ at $t=1$; after that, quotienting by $G$ means that the remaining gauge transformations are smooth at $t=1$). Using this description of  $\tilde{M}_n(c)$ we can finally show
\begin{proposition} $\tilde{M}_n(c)$  equipped with the metric (\ref{smetric}) is a smooth hyperk\"ahler manifold. The tangent space at a solution $(T_0,T_1,T_2,T_3)$ is described by the equations (\ref{tangent}).\label{smooth2} \end{proposition}
\begin{pf} 
Since the metric (\ref{smetric}) may be degenerate, we still have to show that the moment map equations on $N_n\times M_{U(n)}(c-1)$ are everywhere transversal. Consider a particular point in $M_{U(n)}(c-1)$ which we represent by a solution $m=(T_0,T_1,T_2,T_3)$  with $T_0(+\infty)=0$ and $T_i(+\infty)=\tau_i$, $i=1,2,3$. Let $\mu$ be the hyperk\"ahler moment map for the action of $G$ on $N_n\times M_{U(n)}$. We observe that the image of $d\mu_{|_m}$ contains the image of $d\mu^\prime_{|_m}$, $\mu^\prime$ being the hyperk\"ahler moment map for the action of $G$ on $N_n\times M(\tau_1,\tau_2,\tau_3)$ (Kronheimer's definition of $M(\tau_1,\tau_2,\tau_3)$ was recalled in  the previous section). The metric on $N_n\times M(\tau_1,\tau_2,\tau_3)$ is non-degenerate and, as $G$ acts freely, $d\mu^\prime_{|_m}$ is surjective. Thus $d\mu$ is surjective at each point in $N_n\times M_{U(n)}(c-1)$ and $\tilde{M}_n(c)$ is smooth. \end{pf}

We observe that, as in the case of $M_{U(n)}(c)$, $\tilde{M}_n(c)$ has isometric actions of the torus $T^n$ (defined as the diagonal subgroup of $U(n)$), of the symmetric group $S_n$, and of $SU(2)$. In particular, the hyperk\"ahler moment map for the action of $T^n$ is still given by the values of $T_1,T_2,T_3$ at infinity (cf. Proposition \ref{moment}).
\par
 We can describe the topology of $\tilde{M}_n(c)$:
\begin{proposition} $\tilde{M}_n(c)$ is a principal $T^n$-bundle over the configuration space $\tilde{C}_n({\Bbb R}^3)$ of $n$ distinct points in ${\Bbb R}^3$.\label{topology}\end{proposition}
We postpone identifying this bundle until the next section (Proposition \ref{bundle}).
\begin{pf} The space $\tilde{C}_n({\Bbb R}^3)$ is the space of regular triples in the subalgebra of diagonal matrices and the moment map $\mu$ for the action of $T^n$ gives us a projection $\tilde{M}_n(c)\rightarrow \tilde{C}_n({\Bbb R}^3)$.
Let us consider a fixed regular triple $(\tau_1,\tau_2,\tau_3)$ and all elements of $\tilde{M}_n(c)$ with $T_i(+\infty)=\tau_i$, $i=1,2,3$, i.e. $\mu^{-1}(\tau_1,\tau_2,\tau_3)$. For each such solution we can make $T_0$ identically $0$ by some gauge transformation $g$ with $g(0)=1$. This is defined uniquely up to the action of ${\cal G}\times T^n$ and so the set of $T^n$-orbits projecting via $\mu$ to $(\tau_1,\tau_2,\tau_3)$  can be identified with the set of solutions to Nahm's equations with $T_0\equiv 0$, $T_1,T_2,T_3$ having the appropriate residues at $t=0$ and being conjugate to $\tau_1,\tau_2,\tau_3$ at infinity. By the considerations at the beginning of this section this space is the hyperk\"ahler quotient of $N_n\times M(\tau_1,\tau_2,\tau_3)$ by $U(n)$. The arguments of \cite{BielJLMS} show that the corresponding complex-symplectic quotient can be identified with the intersection of a regular semisimple adjoint orbit of $GL(n,{\Bbb C})$ with the slice to the regular nilpotent orbit. This intersection is a single point. Finally, in order to identify in this case the hyperk\"ahler quotient with the complex-symplectic one we can adapt the argument in the proof of Proposition 2.20 in \cite{Hur}.\end{pf}  
Our next task is to describe the complex structure of $\tilde{M}_n(c)$ (because of the action of $SU(2)$ all complex structures are equivalent). As usual (cf. \cite{Don}), if we choose a complex structure, say $I$, we can introduce complex coordinates on the moduli space of solutions to Nahm's equations by writing $\alpha=T_0+iT_1$ and $\beta=T_2+iT_3$. The Nahm equations can be then written as one complex and one real equation:
\begin{eqnarray} & &\frac{d\beta}{dt} = [\beta,\alpha]\label{complex}\\
 & &\frac{d\,}{dt}(\alpha+\alpha^\ast) =[\alpha^\ast,\alpha]+[\beta^\ast,\beta].\label{real}\end{eqnarray}
By the remark made at the beginning of this section, $\tilde{M}_n(c)$ is the hyperk\"ahler quotient of the product manifold $N_n\times M_{U(n)}(c-1)$. We shall show that as a complex symplectic manifold $\tilde{M}_n(c)$ is the complex-symplectic quotient of $N_n\times M_{U(n)}(c-1)$. Let us recall the complex structure of $N_n$ \cite{Don,Hurt,BielJLMS,Dan2}. Let $e_1,\ldots,e_n$ denote the standard basis of ${\Bbb C}^n$. There is a unique solution $w_1$ of the equation
\begin{equation} \frac{dw}{dt}=-\alpha w\label{alphato0}\end{equation}
with
\begin{equation} \lim_{t\rightarrow 0}\left(t^{-(n-1)/2}w_1(t)-e_1\right)=0\label{w1}.\end{equation}
Setting $w_i(t)=\beta^{i-1}(t)w_1(t)$, we obtain a solution to \eqref{alphato0} with 
$$ \lim_{t\rightarrow 0}\left(t^{i-(n+1)/2}w_i(t)-e_i\right)=0.$$
The complex gauge transformation $g(t)$ with $g^{-1}=(w_1,\ldots,w_n)$ makes $\alpha$ identically zero and sends $\beta(t)$ to the constant matrix
\begin{equation} B(\beta_1,\ldots,\beta_n)=\begin{pmatrix} 0 & \ldots & 0 & (-1)^{n+1} S_n \\ 1 & \ddots & & (-1)^n S_{n-1}\\ & \ddots & \ddots & \vdots \\ 0&\ldots & 1 & S_1 \end{pmatrix}.\label{betaconst}\end{equation}
Here $\beta_i$ denote the (constant) eigenvalues of $\beta(t)$ and $S_i$ is the $i$-th elementary symmetric polynomial in $\{\beta_1,\ldots,\beta_n\}$.\newline
The mapping $(\alpha,\beta)\rightarrow (g(1),B)$ gives a biholomorphism between $(N_n,I)$ and $Gl(n,{\Bbb C})\times {\Bbb C}^n$ \cite{BielJLMS}.

We describe the complex structure of $\tilde{M}_n(c)$ as follows:
\begin{proposition} There exists a $T^n$-equivariant biholomorphism between $\tilde{M}_n(c)$ and an open subset of 
$$\left(\coprod_{\frak n}\bigl\{[g,b]\in Gl(n,{\Bbb C})\times_N({\frak d}+{\frak n}); gbg^{-1}\enskip \text{is of the form \eqref{betaconst}}\bigr\}\right)\biggl/\sim ,$$
 where ${\frak d}$ denotes diagonal matrices, the union is over unipotent algebras ${\frak n}$  (with respect to ${\frak d}$) and $N=\exp {\frak n}$. Furthermore, the relation $\sim$ is given as follows: $[g,d+n] \sim [g^\prime,d^\prime+n^\prime]$ if and only if  $n\in {\frak n},n^\prime \in {\frak n}^\prime$,  and either ${\frak n}^\prime\subset {\frak n}$ and there exists an $m\in N$ such that $gm^{-1}=g^\prime,\Ad(m)(d+{\frak n})=d^\prime+{\frak n}^\prime$ or vice versa (i.e. ${\frak n}\subset {\frak n}^\prime$ etc.).  \label{Mascomplex}\end{proposition}
\begin{remarknon} It will follow from the description of the twistor space that  this  biholomorphism is actually onto. Proving this right now would require showing that the $T^n$-action on $\tilde{M}_n(c)$ extends to the global action of $\bigl({\Bbb C}^\ast\bigr)^n$. This, in turn, requires showing existence of solutions to a mixed Dirichlet-Robin problem on the half-line - something that seems quite tricky.\end{remarknon}
\begin{pf}
Fix a unipotent algebra ${\frak n}$ and consider the set of all solutions $(\alpha,\beta)=(T_0+iT_1,T_2+iT_3)$ on $[1,+\infty)$ such that the intersection of the sum of positive eigenvalues of $\text{ad}(iT_1(+\infty))$ with $C(\beta(+\infty))$ is contained in ${\frak n}$. Let $M({\frak n};c-1)$ be the corresponding subset of $M_{U(n)}(c)$. We observe that, since $(T_1(+\infty),T_2(+\infty),T_3(+\infty))$ is a regular triple, the projection of $T_1(+\infty)$ onto ${\frak d}^{\Bbb C}\cap C(\beta(+\infty))$ is a regular element, and so ${\frak n}$ contains the unipotent radical of a Borel subalgebra of $C(\beta(+\infty))$ for any element of $M({\frak n};c-1)$. Using gauge freedom, we always make $T_0(+\infty)=0$ and, by Proposition 4.1 of Biquard \cite{Biq}, such a representative is of the form $g\bigl(\alpha(+\infty), \beta(+\infty)+\text{Ad}(\exp\{-\alpha(+\infty)t\})n\bigr)$, where $n\in {\frak n}$ and $g$ is a bounded $Gl(n,{\Bbb C})$-valued gauge transformation. The transformation $g$ is defined modulo $\exp\{-\alpha(+\infty)t\}g_0\exp\{\alpha(+\infty)t\}$ with $g_0\in P=\exp({\frak d}+{\frak n})$. Since $T_0(+\infty)=0$ and $T_0$ is decaying exponentially fast, $g$ has a limit (in $T^{\Bbb C}$) at $+\infty$. If we replace $g(t)$ by $g^\prime(t)=g(t)g(+\infty)^{-1}\exp\{-\alpha(+\infty)t+c\alpha(+\infty)\}$, then $(\alpha,\beta)=g^\prime(0,\beta(+\infty)+n^\prime)$ for an $n^\prime\in {\frak n}$. The transformation $g^\prime$, which satisfies (at infinity) the boundary condition of an element of ${\cal G}(c-1)^{\Bbb C}$, is now defined modulo constant gauge transformations in $N$. Moreover $g^\prime(1)$ is independent of ${\cal G}(c-1)$ and we obtain a map $\phi:M({\frak n})\rightarrow Gl(n,{\Bbb C})\times_N ({\frak d}+{\frak n})$ by sending $(\alpha,\beta)$ to $(g^\prime(1),\beta(+\infty)+n^\prime)$. Considering the infinitesimal version of this construction shows that $\phi$ is holomorphic.
\par
Since $\phi$ is $U(n)$-equivariant, it is (locally) $Gl(n,{\Bbb C})$-equivariant. We can adapt the argument of Proposition 2.20 in \cite{Hurt} to show that $\tilde{M}_n(c)$ is the complex-symplectic quotient of $N_n\times M_{U(n)}(c-1)$ by (local action of) $Gl(n,{\Bbb C})$. Let us restrict attention to $N_n\times M({\frak n})$. The complex symplectic moment map at the point $(g,B)$ of $N_n$ is $-g^{-1}Bg$ (here $g\in Gl(n,{\Bbb C})$ and $B$ is of the form \eqref{betaconst})  and the complex symplectic moment map at the point corresponding to $[g^\prime,\beta_d+n]$ is $g^\prime(\beta_d+n)g^{\prime-1}$ (here $\beta_d$ is diagonal and $n\in{\frak n}$). The moment map equation for the diagonal action of $Gl(n,{\Bbb C})$ is $g^{-1}Bg = g^\prime(\beta_d+n)g^{\prime-1}$. If we now quotient by $Gl(n,{\Bbb C})$, i.e. send $g$ to identity, we shall end up with the set of $[g^\prime,b]\in Gl(n,{\Bbb C})\times_N({\frak d}+{\frak n})$ such that $g^\prime b g^{\prime-1}=B$ ($B$ is determined by the diagonal part of $b$). This identifies the charts described in this proposition. By going through the procedure we can conclude that the charts for different ${\frak n}$ are matched as claimed.
\par
So far we have shown that there is a holomorphic map $\phi$ from $\tilde{M}_n(c)$ to the manifold $M$ described in the statement. We still have to show that $\phi$ is 1-1. By construction our map is $T^n$-equivariant, and so $\bigl({\Bbb C}^\ast\bigr)^n$-equivariant (where the action is defined). Since the $\bigl({\Bbb C}^\ast\bigr)^n$-action on $M$ is free, it is free on $\tilde{M}_n(c)$. Furthermore the $\bigl({\Bbb C}^\ast\bigr)^n$-action on $M$ leaves invariant sets of the form $M\cap \bigl(Gl(n,{\Bbb C})\times_N (d+{\frak n})\bigr)$, $d\in {\frak d}$. Each such set is single orbit of $\bigl({\Bbb C}^\ast\bigr)^n$ and so $\phi$ is 1-1. 
\end{pf} 
The above description of $\tilde{M}_n(c)$ is rather complicated. We remark that the open dense subset where $\beta(+\infty)$ is regular corresponds to ${\frak n}=0$, i.e. to $$\{(\beta_d,g); \beta_d=\diag (\beta_1,\dots,\beta_n), \beta_i\neq\beta_j\enskip\text{if} \enskip i\neq j,\; g\beta_dg^{-1}=B(\beta_1,\dots,\beta_n)\}.$$ 
We shall denote the corresponding subset of $\tilde{M}_n(c)$ by $\tilde{M}_n^{\rm reg}(c)$. We observe that an element $g$ of $Gl(n,{\Bbb C})$ which sends $\diag (\beta_1,\dots,\beta_n)$ to $B(\beta_1,\dots,\beta_n)$ is of the form 
\begin{equation} g=V(\beta_1,\ldots,\beta_n)^{-1}\diag (u_1,\ldots,u_n)\label{Vu}\end{equation}
 where $u_i\neq 0$ and $V(\beta_1,\ldots,\beta_n)$ is the Vandermonde matrix, i.e. $V_{ij}=(\beta_i)^{j-1}$. We can calculate the complex symplectic form $\omega=\omega_2+i\omega_3$ on $\tilde{M}_n^{\rm reg}(c)$:
\begin{proposition} The complex symplectic form $\omega$ on $\tilde{M}_n^{\rm reg}(c)$ is given, in coordinates $\beta_i,u_i$, $i=1,\ldots,n$, by
\begin{equation} \sum_{i=1}^n \frac{du_i}{u_i}\wedge  d\beta_i-\sum_{i<j}\frac{d\beta_i\wedge d\beta_j}{\beta_i-\beta_j} .\label{omegaf}\end{equation}
\label{complexform}\end{proposition}
\begin{pf} First, we calculate $\omega$ on the subset of $M_{U(n)}(c-1)$ where $\beta(+\infty)$ is regular. This subset is biholomorphic to $Gl(n,{\Bbb C})\times\{\text{\em regular elements of ${\frak h}^{\Bbb C}$}\}$ and according to the proof of Proposition \ref{Mascomplex}, an element $(\alpha,\beta)$ of this set corresponding to $(g,\beta_d)\in Gl(n,{\Bbb C})\times{\frak h}$ can be written as $(\alpha,\beta)=(-\dot{g}(t)g^{-1},g(t)\beta_dg(t)^{-1})$, where $g(t)$ is a complex gauge transformation with $g(0)=g$. Therefore a tangent vector $(a(t),b(t))$ at $(\alpha,\beta)$ can be written as 
\begin{equation} (a,b)=\bigl(-g\dot{\rho}g^{-1},g\bigl(b_d+[\rho,\beta_d]\bigr)g^{-1}\bigr),\label{a,b}\end{equation}
where $\rho$ is dual to $g^{-1}dg$ and $b_d$ is dual to $d\beta_d$. The complex symplectic form on $M_{U(n)}(c-1)$ is given by
$$\omega=(c-1)\tr\bigl(d\alpha(+\infty)\wedge d\beta(+\infty)\bigr)+\int_0^{+\infty}\tr\bigl(d\alpha\wedge d\beta-d\alpha(+\infty)\wedge d\beta(+\infty)\bigr).$$
For two tangent vectors $(a,b)$ and $(\hat{a},\hat{b})$, corresponding, via \eqref{a,b}, to $(\rho,b_d)$ and $(\hat{\rho},\hat{b}_d)$ we obtain
$$\omega=-\tr\bigl(b_d\hat{\rho}-\rho\hat{b}_d-[\rho,\hat{\rho}]\beta_d\bigr),$$
where $\rho=\rho(0),\hat{\rho}=\hat{\rho}(0)$. To calculate the symplectic form on $\tilde{M}_n^{\rm reg}(c)$ it remains to substitute \eqref{Vu} for $g$. Let us write $u$ for $\diag(u_1,\ldots, u_n)$. Then $\rho$ becomes dual to $u^{-1}du-u^{-1}dVV^{-1}u$. Let us write $\nu$ for the tangent vector dual to $u^{-1}du$ and $\Upsilon$ for the tangent vector dual to $dVV^{-1}$. Since $\nu$ is diagonal and the $i$-th row of $\Upsilon$ is of the form $b_is$ (here we write $b_d=\diag(b_1,\ldots,b_n)$), for a covector $s$, we can write $\omega$ as
$$\omega=-\tr\bigl(b_d\hat{\nu}-\nu\hat{b}_d-[\Upsilon,\hat{\Upsilon}]\beta_d\bigr).$$ 
It remains to calculate $\tr[\Upsilon,\hat{\Upsilon}]\beta_d$. Let us write $W_{ij}$ for the $(i,j)$-th entry of $V^{-1}$, i.e.
\begin{equation} W_{ij}=(-1)^{n-i}
 S_{n-i}(\beta_1,\dots,\hat{\beta}_j,\dots,\beta_n)\Big/\prod_{k\neq j}(\beta_j-\beta_k),\label{Wij}\end{equation}
 $S_k$ being the $k$-th elementary symmetric polynomial ($S_0=1$). We calculate the $(i,i)$-th entry of $[\Upsilon,\hat{\Upsilon}]$ as 
$$\sum_j (b_i\hat{b}_j-\hat{b}_ib_j)\left(\sum_k (k-1)\beta_i^{k-2}W_{kj}\right) \left(\sum_k (k-1)\beta_j^{k-2}W_{ki}\right).$$
This means that
$$\tr[\Upsilon,\hat{\Upsilon}]\beta_d=\sum_{i<j}(b_i\hat{b}_j-\hat{b}_ib_j) (\beta_i-\beta_j)\left(\sum_k (k-1)\beta_i^{k-2}W_{kj}\right) \left(\sum_k (k-1)\beta_j^{k-2}W_{ki}\right).$$
Formula \eqref{omegaf} will be proven if we can show (for $i\neq j$)
the following identity:
\begin{equation} \left(\sum_k (k-1)\beta_i^{k-2}W_{kj}\right) \left(\sum_k (k-1)\beta_j^{k-2}W_{ki}\right)=\frac{-1}{(\beta_i-\beta_j)^2}.\label{identity} \end{equation}
According to \eqref{Wij} we have
\begin{equation}\sum_k (k-1)\beta_i^{k-2}W_{kj}=\frac{\sum_k (k-1)\beta_i^{k-2} (-1)^{n-k}S_{n-k}(\beta_1,\dots,\hat{\beta}_j,\dots,\beta_n)}{\prod_{s\neq j}(\beta_j-\beta_s)}.\label{identity2}\end{equation}
We compute the numerator of this expression. We set $p=n-1$ and $(a_1,\ldots,a_p)=(\beta_1,\dots,\hat{\beta}_j,\dots,\beta_n)$. Then the numerator can be written as
$$\sum_{s=0}^p (p-s)(-1)^s a_i^{p-1-s} S_s(a_1,\ldots,a_p)=\frac{d\,}{dt}\left(\sum_{s=0}^p (-1)^sS_s t^{p-s}\right)_{t=a_i}.$$
Since $\sum S_s t^s=\prod(1+a_s t)$, we can rewrite the expression under the derivative as
$$\sum_{s=0}^p (-1)^sS_s t^{p-s}=\prod_{s=0}^p (t-a_s).$$
Taking the derivative and substituting $a_i$ for $t$, finally gives
$$\sum_{s=0}^p (p-s)(-1)^s a_i^{p-1-s}=\prod_{s\neq i}(a_i-a_s).$$
Going back to \eqref{identity2}, we have
$$\sum_k (k-1)\beta_i^{k-2}W_{kj}=\frac{\prod_{s\neq i,j}(\beta_i-\beta_s)}{\prod_{s\neq j}(\beta_j-\beta_s)},$$
from which \eqref{identity} follows.
\end{pf}

\begin{remark} Setting $$p_i=u_i\Big/\prod_{j>i}(\beta_i-\beta_j),$$
the formula \eqref{omegaf} can be rewritten as
$$\omega=\sum_{i=1}^n \frac{dp_i}{p_i}\wedge d\beta_i.$$
\label{newform}\end{remark}

\section{The twistor space and the metric on $\tilde{M}_n(c)$\label{three}}

We shall now identify the twistor space $Z(c)$ of $\tilde{M}_n(c)$. As a first step, we observe, after Hitchin et al.\ \cite{HKLR}, that the hyperk\"ahler moment map $\mu$ for the $T^n$-action defines a moment map, also denoted by $\mu$, for the complex-symplectic form along the fibers $Z(c)\rightarrow {\Bbb C}P^1$. This $\mu$ is a map from $Z(c)$ to $O(2)\otimes {\Bbb C}^n$. 
 We shall first identify the open subset $Z^{\rm reg}(c)$ of $Z(c)$ defined as the set  
\begin{equation} Z^{\rm reg}(c)=\mu^{-1}\left(O(2)\otimes {\Bbb C}^n - O(2)\otimes \Delta\right),\label{reg}\end{equation} 
where $\Delta$ is the generalized diagonal in ${\Bbb C}^n$.    
In terms of the coordinates $(\beta_1,\dots,\beta_n)$ and $(u_1,\ldots,u_n)$ given by \eqref{Vu}, $Z^{\rm reg}(c)$ has the following description:
\begin{proposition}  $Z^{\rm reg}(c)$ is obtained by taking two copies of ${\Bbb C}\times ({\Bbb C}^n-\Delta)\times ({\Bbb C}^\ast)^n$ with coordinates $(\zeta,\beta_i,u_i)$ and $(\tilde{\zeta},\tilde{\beta}_i,\tilde{u}_i)$, $i=1,\dots,n$, and identifying over $\zeta\neq 0$ by
$$\begin{array}{l} \tilde{\zeta}=\zeta^{-1}\\
 \tilde{\beta}_i=\zeta^{-2}\beta_i\\ \tilde{u}_i=\zeta^{-(n-1)}\exp\{-c\beta_i/\zeta\}u_i.\end{array}$$
The real structure is given by 
$$\begin{array}{l} \zeta\mapsto -1/\bar{\zeta}\\ \beta_i\mapsto -\bar{\beta}_i/\bar{\zeta}^2\\ 
u_i\mapsto \bar{u}_i^{-1}\left(1/\bar{\zeta}\right)^{n-1}\prod_{j\neq i}(\bar{\beta}_i-\bar{\beta}_j)e^{c\bar{\beta}_i/\bar{\zeta}}.\end{array}$$ 
Finally, the complex symplectic form along the fibers is given by \eqref{omegaf}.
\label{twistor1}\end{proposition}
\begin{pf}
For any hyperk\"ahler moduli space of solutions to Nahm's equations one can trivialize the twistor space  by choosing an affine coordinate $\zeta$ on ${\Bbb C}P^1$ and then putting $\eta=\beta+(\alpha+\alpha^\ast)\zeta-\beta^\ast\zeta^2$, $u=\alpha-\beta^\ast\zeta$ for $\zeta\neq\infty$, and $\tilde{\eta}=\beta/\zeta^{2}+(\alpha+\alpha^\ast)/\zeta-\beta^\ast$, $\tilde{u}=-\alpha^\ast-\beta/\zeta$ for $\zeta\neq 0$. Then, over $\zeta\neq 0,\infty$, we have $\tilde{\eta}=\eta/\zeta^2$, $\tilde{u}=u-\eta/\zeta$. Moreover, the real structure is $\zeta\mapsto -1/\bar{\zeta}$, $\eta\mapsto -\eta^\ast/\bar{\zeta}^2$, $u\mapsto -u^\ast+\eta^\ast/\bar{\zeta}$ (cf. \cite{Dan2,Biq2}. 
\par
We now have to go through the procedure in the proof of Proposition \ref{Mascomplex} to describe $Z^{\rm reg}$ in coordinates $(\zeta,\beta_i,u_i)$ and $(\tilde{\zeta},\tilde{\beta}_i,\tilde{u}_i)$. First we describe the twistor space of $N_n$ in coordinates $(g,B)$ and $(\tilde{g},\tilde{B})$ defined right after \eqref{betaconst} (cf. \cite{Dan2}).
Going through the procedure assigning $(g,B)$ to $(\alpha,\beta)$ we see that $\tilde{B}=B\bigl(\beta_1/\zeta^{-2},\dots,\beta_n/\zeta^{-2}\bigr)$. On the other hand $g$ is given by $g=g(1)$ where $g(t)$ is a complex gauge transformation such that $\frac{d}{dt}g^{-1}=-u g^{-1}$. This means that $g(t)$ makes $u$ identically zero. We observe that $\exp\{-Bt/\zeta\}g(t)$ makes $\tilde{u}$ identically zero and $\tilde{\eta}$ into $B/\zeta^2$. The initial value for the solution $g^{-1}$ depends on $\zeta$ and so we can write $\tilde{g}(t)=U\exp\{-Bt/\zeta\}g(t)$ for some constant matrix $U$. If we are to get the form \eqref{betaconst}, we must have $U=U^\prime d(\zeta)$, where 
\begin{equation}d(\zeta)=\diag \left(\zeta^{-(n-1)}, \zeta^{-(n-3)},\dots, \zeta^{n-1}\right).\label{dzeta}\end{equation}
In addition $U^\prime$ commutes with $B\bigl(\beta_1/\zeta^{-2},\dots,\beta_n/\zeta^{-2}\bigr)$. Moreover, the initial value for the equation $\frac{d}{dt}g^{-1}=-\alpha g^{-1}$ depends only on the residues of $u,\eta,\tilde{u},\tilde{\eta}$ and therefore $U^\prime$ does not depend on $B$. Since the initial values belong to $SU(n)$, we also have $U^\prime\in SU(n)$. It follows that $U^\prime$ belongs to the center of $SU(n)$. This is only an ambiguity in the choice of trivialization and it does not affect the twistor space.
Similar considerations show that the real structure sends $B(\beta_1,\dots,\beta_n)$  to $B\bigl(-\bar{\beta}_1/\bar{\zeta}^{-2},\dots,-\bar{\beta}_n/\bar{\zeta}^{-2}\bigr)$ and $g$ to $r(\zeta)\exp\{B^\ast/\bar{\zeta}\}\left(g^\ast\right)^{-1}$ where 
$$ r_{ij}(\zeta)=\left\{\begin{matrix} 0 & \text{if} & i+j\neq n+1\\
                  (-1)^{j-1}\bar{\zeta}^{n+1-2j} & \text{if} & i+j= n+1\end{matrix}\right..$$
This time the remaining ambiguity is given by a real element in the center of $SU(n)$, i.e. $-1$ if $n$ is even.
\par
We now go through a similar procedure for the subset of $M_{U(n)}(c-1)$ where $\beta(+\infty)$ is regular. We have assigned in the proof of Proposition \ref{Mascomplex} to each element of this set a pair $(g,\beta(+\infty)$. We already know how $\beta(+\infty)$ changes (as it is given by the complex moment map for a torus action). The proof of Proposition \ref{Mascomplex} shows that the other coordinates, $g$ on $\{\zeta\neq\infty\}$ and $\tilde{g}$ on $\{\zeta\neq 0\}$, are related by $\tilde{g}=g\exp\{-(c-1)\beta(+\infty)/\zeta\}$. The real structure sends $g$ to $\left(g^\ast\right)^{-1}\exp\{(c-1)\beta(+\infty)^\ast/\bar{\zeta}\}$.
\par
Finally we have to go to the complex-symplectic quotient as in the proof of Proposition \ref{Mascomplex}. We end up with $(g,\beta_d)$ and $(\tilde{g},\tilde{\beta}_d)$ where $\beta_d=\diag(\beta_1,\ldots, \beta_n)$ and $g\beta_dg^{-1}=B(\beta_1,\ldots,\beta_n)$ (and similarily for $(\tilde{g},\tilde{\beta}_d)$). We see that $\beta_i$ and $\tilde{\beta}_i$ are related as stated and $\tilde{g}= d(\zeta)\exp\{-B/\zeta\}g\exp\{-(c-1)\beta_d/\zeta\}$. Since $\exp\{-B/\zeta\}g=g\exp\{-\beta_d/\zeta\}$, $\tilde{g}= d(\zeta)g\exp\{-c\beta_d/\zeta\}$. If we now go to the coordinates $u_i,\tilde{u}_i$ defined by \eqref{Vu}, we see that they change as required, since the $(i,j)$-th entry of $V^{-1}$ is given by \eqref{Wij} and the $\beta_i$ change as prescribed (i.e. as sections of $O(2)$). A similar argument shows that the real structure is, up to a sign, the one described in the statement (it is enough to compare the last row in $r(\zeta)\bigl(V^{-1}\diag \{u_i\}\bigr)^{\ast-1}\diag\bigl\{e^{c\bar{\beta}_i/\bar{\zeta}}\bigr\}$ and in $V^{-1}\bigl(-\bar{\beta}_1/\bar{\zeta}^{-2},\dots,-\bar{\beta}_n/\bar{\zeta}^{-2}\bigr)\diag \{u^\prime_i\}$). We shall see shortly (Proposition \ref{twistor2}) that the negative of the real structure described in the statement does not admit any sections (a section would be equivalent to a complex number with imaginary modulus).
The formula for the complex symplectic structure is a direct consequence of Proposition \ref{complexform}. \end{pf}

We now wish to find the full twistor space and the metric on $\tilde{M}_n(c)$ and this means finding a family of real sections. We know their projections to $O(2)\otimes {\Bbb C}^n$: they are given by $\left(\beta+(\alpha+\alpha^\ast)\zeta-\beta^\ast\right)(+\infty)$ (cf. \cite{HKLR}) and are parameterized by $n$ distinct points in ${\Bbb R}^3$ with coordinates $(x_i,{\rm Re}\,z_i,$ ${\rm Im}\,z_i)$, $i=1,\ldots,n$, where $x_i=\sqrt{-1}T_1(+\infty),z_i=\beta(+\infty)$. In other words we have $n$  curves $S_i=\{(\zeta,\eta);\eta=z_i+2x_i\zeta-\bar{z}_i\zeta^2\}$ in $T{\Bbb C}P^1$ (here $\eta$ is the fiber coordinate). According to Proposition \ref{twistor1} the $u_i$ coordinate of a real section of $Z(c)$  changes as a non-zero section of the bundle $L^{c}(k-1)$ (with the transition function  $\zeta^{k-1}e^{c\eta/\zeta}$ from $\infty$ to $0$) over $S_i$. This is true only away from the intersection points of the curves $S_i$ and we have to understand what happens to the section at these points.  Two curves $S_i=\{(\zeta,\eta);\eta=z_i+2x_i\zeta-\bar{z}_i\zeta^2\}$ and $S_j=\{(\zeta,\eta);\eta=z_j+2x_j\zeta-\bar{z}_j\zeta^2\}$  intersect in a pair of distinct points $a_{ij}$ and $a_{ji}$, where
\begin{equation} a_{ij}=\frac{(x_i-x_j)+r_{ij}}{\bar{z}_i-\bar{z}_j}, \quad r_{ij}=\sqrt{(x_i-x_j)^2 +|z_i-z_j|^2}.\label{aij}\end{equation}
We have:
\begin{proposition} The real sections of the twistor space $Z(c)$ of $\tilde{M}_n(c)$ are given, over $\zeta\neq \infty$, by $\bigl(\beta_1(\zeta), \ldots,\beta_n(\zeta),u_1(\zeta),\ldots,u_n(\zeta)\bigr)$, where
\begin{align*} & \beta_i(\zeta)=z_i+2x_i\zeta-\bar{z}_i\zeta^2,\\
&  u_i(\zeta)=A_i\prod_{j\neq i}(\zeta-a_{ji})e^{c(x_i-\bar{z}_i\zeta)},\end{align*}
where $(x_i,z_i)$, $i=1,\ldots,n$, are distinct points in ${\Bbb R}\times{\Bbb C}$ and $A_i$ are complex numbers satisfying
$$ A_i\bar{A}_i=\prod_{j\neq i}\bigl(x_i-x_j+r_{ij}\bigr).$$
\label{twistor2}\end{proposition}
\begin{remarknon} Given Proposition \ref{topology}, this finally shows that the biholomorphism of Proposition \ref{Mascomplex} is onto.\end{remarknon}
\begin{pf}
Consider a real section $s$ of $Z(c)$ (corresponding to a solution $(T_0,T_1,T_2,T_3)$) which projects to a given real section $(\beta_1(\zeta),\ldots,\beta_n(\zeta))$ of $O(2)\otimes{\Bbb C}^n$. For a generic section the intersection points of the $\beta_s$ are all distinct. We consider the point $a_{ji}$ at which $\beta_i$ intersects $\beta_j$ and let us assume that no other $\beta_s$ intersect there. We recall that $\sqrt{-1}T_1(\zeta)=\frac{1}{2}(\alpha+\alpha^\ast)-\beta^\ast\zeta$ and, hence, $\sqrt{-1}T_1(\zeta)(+\infty)_{ss}=x_s-\bar{z}_s\zeta$. This means that $\sqrt{-1}T_1(a_{ji})(+\infty)_{jj}<\sqrt{-1}T_1(a_{ji})(+\infty)_{ii}$, and so, with respect to the complex structure corresponding to $a_{ji}\in {\Bbb C}P^1$, the solution $(T_0,T_1,T_2,T_3)$ belongs to the chart described in Proposition \ref{Mascomplex} with ${\frak n}$ generated by the matrix with the only non-zero entry having coordinates $(i,j)$.   Let us write $s$ as $(\beta_i(\zeta),u_i(\zeta))$, $i=1,\ldots,n$, in a neighbourhood of $a_{ji}$, $\zeta\neq a_{ji}$ (notice that the procedure of Proposition \ref{twistor1} does assign well-defined complex numbers $u_1(\zeta),\ldots,u_n(\zeta)$ to each $\zeta\neq a_{ji}$).  
According to the proof of Proposition \ref{Mascomplex} there is an element $m(\zeta)\in N=\exp{\frak n}$ such that the following expression
$$V\bigl(\beta_1(\zeta),\ldots,\beta_n(\zeta)\bigr)^{-1} \diag\bigl(u_1(\zeta),\dots,u_n(\zeta)\bigr)
m(\zeta)$$
has an invertible limit at $\zeta=a_{ji}$.   
Let $W_{kl}(\zeta)$ denotes the $(k,l)$-th entry of $V\bigl(\beta_1(\zeta),\ldots,\beta_n(\zeta)\bigr)^{-1}$ and let $p(\zeta)$ denote the only non-zero non-diagonal entry of $\diag\bigl(u_1(\zeta),\dots,u_n(\zeta)\bigr)m(\zeta)$ ($p(\zeta)$ is the $(i,j)$-th entry). We then have that $W_{kj}u_j +W_{ki}p$ and $W_{ki}u_i$ have a finite limit at $\zeta=a_{ji}$, for all $k=1,\ldots,n$. From the formula \eqref{Wij} a finite limit for $W_{ni}u_i$ implies that $u_i(a_{ji})=0$, while the nonvanishing of the last row of $V^{-1}\diag(u_s)m$ means that $a_{ji}$ is a single zero of $u_i$.      
If more than two sections $\beta_s(\zeta)$ meet at $a_{ji}$ the considerations are similar but involve larger ${\frak n}$. We can conclude the $a_{ji}$ contribute precisely $n-1$ zeros of $u_i$ (counting multiplicities) and, given Proposition \ref{twistor1}, this proves the formula for $u_i(\zeta)$ as soon as we show that $u_i$ has no other zeros, or, equivalently, no poles. To prove this latter statement it is enough to show that $u_j$ does not have a pole at $a_{ji}$. We go back to the situation when ${\frak n}$ is one-dimensional, and where we concluded that  $W_{kj}u_j +W_{ki}p$ has a finite limit at $\zeta=a_{ji}$ for all $k=1,\ldots,n$. We can write $W_{nj}u_j +W_{ni}p$ as $(fu_j+gp)/(\beta_i-\beta_j)$ where $f$ and $g$ have finite limits at $\zeta=a_{ji}$. We then have
$$W_{n-1,j}u_j +W_{n-1,i}p=-\left(fu_j\bigl(\sum_{s\neq j}\beta_s\bigr)+gp\bigl(\sum_{s\neq i}\beta_s\bigr)\right)\frac{1}{\beta_i-\beta_j}$$
which can be rewritten as
$$-fu_j-\bigl(\sum_{s\neq i}\beta_s\bigr)\frac{fu_j+gp}{\beta_i-\beta_j}.$$
Since the second term has a finite limit, so does $fu_j$ and hence $u_j$. Again, if more than two sections $\beta_s(\zeta)$ meet at $a_{ji}$ the considerations are similar but involve larger ${\frak n}$. Thus we have shown the second formula of the statement. The last formula follows from the reality condition and the fact that $a_{ji}=-1/\bar{a}_{ij}$ (this calculation also eliminates the $\pm 1$ ambiguity in the choice of the real structure in the proof of \ref{twistor1}).\end{pf}

We can finally identify $\tilde{M}_n(c)$ as a $T^n$-bundle over the configuration space $\tilde{C}_n({\Bbb R}^3)$ of $n$ distinct points ${\bf x}_i$ in ${\Bbb R}^3$.
\begin{proposition}  $\tilde{M}_n(c)$ is equivalent to the $T^n$-bundle described in Proposition \ref{bundle0}.\label{bundle}\end{proposition}
\begin{pf} From the last formula in Proposition \ref{twistor2} it follows that $A_i\neq 0$ if, for all $j\neq i$, $z_i\neq z_j$ or $x_i>x_j$. On the other hand, if we put
$$A_I=A_i\prod_{j\in I}a_{ji},$$
for any subset $I$ of $\{j;j\neq i\}$,
then we have
$$A_I\bar{A}_I=\prod\begin{Sb}j\neq i\\j\not\in I\end{Sb}\bigl(x_i-x_j+r_{ij}\bigr) \prod_{j\in I}\bigl(x_j-x_i+r_{ij}\bigr).$$
Let us choose sets $I_1,\ldots,I_n$ such that $I_i\subset\{j;j\neq i\}$ and $j\in I_i\Leftrightarrow i\not \in I_j$. Define $U(I_1,\ldots,I_n)$ as the complement of the subset $\bigl\{(x_i,z_i)_{i=1,\dots,n};I_i^c=\{j;z_i=z_j\enskip\text{and} \enskip x_i<x_j\}\bigr\}$ ($I^c_i$ denotes the complement of $I_i$ in $\{j;j\neq i\}$).
The sets $U(I_1,\ldots,I_n)$ cover $\tilde{C}_n({\Bbb R}^3)$ and over each of them the bundle $\tilde{M}_n(c)$ is trivialized by coordinates $\bigl(x_i,z_i,A_{I_i}/|A_{I_i}|\bigr)$. To determine the bundle, choose $i<j$. The bundle restricted to $S_{ij}^2$ is given by the transition function from $U(I_1,\ldots,I_n)$ where $j\not\in I(i)$ to $U(I_1^\prime,\ldots,I_n^\prime)$ where $I_i^\prime=I_i\cup\{j\}$, $I_j^\prime=I_j-\{i\}$, $I_k^\prime=I_k$ for $k\neq i,j$. Let $\phi_k$ be the transition function for the $k$-th generator of $T^n$, i.e. the transition function from $A_{I_k}/|A_{I_k}|$ to $A_{I_k^\prime}/|A_{I_k^\prime}|$. We see that $\phi_k=1$ if $k\neq i,j$, and $\phi_i=a_{ji}/|a_{ji}|$, $\phi_j=|a_{ji}|/a_{ji}$. Therefore $\phi_i=(z_j-z_i)/|z_j-z_i|$ and $\phi_j=\phi_i^{-1}$. It remains to identify the circle bundle over the sphere $x^2+|z|^2=\text{const}$ given by the transition function $z/|z|$ from the region  $U_0=\{z\neq 0\enskip\text{or}\enskip x>0\}$ to the region $U_1=\{z\neq 0\enskip\text{or}\enskip x<0\}$. Let us write the unit $3$-sphere as $\{(u,v)\in{\Bbb C}^2; |u|^2+|v|^2=1\}$. The Hopf bundle is given the $S^1$ action $t\cdot(u,v)=(tu,t^{-1}v)$ and the projection $S^3\rightarrow S^2$ by the map $x=|u|^2-|v|^2$, $z=2uv$. Over $U_0$ this bundle is trivialized by $(x,z,u/|u|)$ and over $U_1$ by $(x,z,|v|/v)$. The transition function is $|z|/z$. Thus $[\phi_i]=-1\in H^1(S^2_{ij},S^1)$.
\end{pf}

We can now calculate the metric on $\tilde{M}_n(c)$. By the remark at the end of section \ref{two}, it is enough to know the metric for $c=-1,0,1$, as the others are obtained by homothety. We shall calculate the metric for $c=1$. The metric for $c=-1$ is the everywhere negative definite version of the Gibbons-Manton metric (this can be seen from the $c=1$ calculation) and the one for $c=0$  is the negative-definite cone metric over a $3$-Sasakian manifold. 

\begin{theorem} $\tilde{M}_n(1)$ is isomorphic, as a hyperk\"ahler manifold, to the Gibbons-Manton manifold $M_{\scriptscriptstyle GM}$ defined in section \ref{zero}.\label{GMmetric}\end{theorem}
\begin{pf} We know from the previous proposition that the two spaces are diffeomorphic. We shall show that the twistor description of  $\tilde{M}_n(1)$ and of the Gibbons-Manton metric coincide. We recall from section \ref{zero} that the latter is a hyperk\"ahler quotient of $M=M_1\times M_2$ by a torus, where $M_1=\bigl(S^1\times {\Bbb R}^3\bigr)^n$ and $M_2={\Bbb H}^{n(n-1)/2}$. With respect to any complex structure $M_1=\bigl({\Bbb C}^\ast\bigr)^n\times {\Bbb C}^n$ and $M_2={\Bbb C}^{n(n-1)/2}\times {\Bbb C}^{n(n-1)/2}$. Let us write the corresponding complex coordinates as $(p_i,\beta_i)$, $i=1,\ldots,n$, on $M_1$ and as $(v_{ij},w_{ij})$, $i<j$, on $M_2$. The complex-symplectic forms corresponding to metrics $g_1$ and $g_2$ are given by
\begin{align} & \sum_{i=1}^n  \frac{dp_i}{p_i}\wedge d\beta_i \label{om1}\\
& \sum_{i<j} dv_{ij}\wedge dw_{ij}.\label{om2}\end{align}
The real sections of the twistor space $Z_1$ of $M_1$ are written, over $\zeta\neq \infty$, as
\begin{equation} \beta_i(\zeta)=z_i+2x_i\zeta-\bar{z}_i\zeta^2,\qquad p_i(\zeta)=B_ie^{x_i-\bar{z}_i\zeta},\label{Z1}\end{equation}
where $B_i\bar{B}_i=1$. The real sections of the twistor space $Z_2$ of $M_2$ are (cf. \cite{Besse}, chapter 13.F):
\begin{equation} v_{ij}(\zeta)=C_{ij}(\zeta-a_{ij}),\qquad w_{ij}(\zeta)=D_{ij}(\zeta-a_{ji}), \label{Z2}\end{equation}
where $a_{ij},a_{ji}$ are roots of $v_{ij}w_{ij}=z_{ij}+2x_{ij}\zeta-\bar{z}_{ij}\zeta^2$ for some $(x_{ij},z_{ij})\in {\Bbb R}\times {\Bbb C}$, i.e.
$$a_{ij}=\frac{x_{ij}+\sqrt{x_{ij}^2+|z_{ij}|^2}}{\bar{z}_{ij}},\qquad a_{ji}=\frac{x_{ij}-\sqrt{x_{ij}^2+|z_{ij}|^2}}{\bar{z}_{ij}}$$
 and 
$$C_{ij}\bar{C}_{ij}=-x_{ij}+\sqrt{x_{ij}^2+|z_{ij}|^2}, \qquad D_{ij}\bar{D}_{ij}=x_{ij}+\sqrt{x_{ij}^2+|z_{ij}|^2}.$$
Here the paricular choice of sections is forced either by the fact the metric is positive definite or by requiring that the $S^1$-action $t\cdot(v_{ij},w_{ij})=(tv_{ij},t^{-1}w_{ij})$ determines the Hopf bundle over the $2$-sphere $x_{ij}^2+|z_{ij}|^2=1$ (this calculation was done in the proof of Proposition \ref{bundle}). 
To obtain the twistor description of the Gibbons-Manton metric we have to perform the complex-symplectic quotient construction along the fibers of $Z_1\oplus Z_2$ with respect to the difference of the forms \eqref{om1} and \eqref{om2}. As in section \ref{zero}, the moment map equations are $v_{ij}w_{ij}=\beta_i-\beta_j$ and so the $a_{ij},a_{ji}$ are given by \eqref{aij}. Since we already know that the manifolds are diffeomorphic, it is sufficient to determine the metric on an open dense subset, e.g. on the set where all $v_{ij}$ are non-zero. Quotienting this set by $\bigl({\Bbb C}^\ast\bigr)^{n(n-1)/2}$ is equivalent to sending  all $v_{ij}$  to $1$. This is achieved by acting by the element $(v_{ij})^{-1}$ of $\bigl({\Bbb C}^\ast\bigr)^{n(n-1)/2}$. By the description of the torus action given in section \ref{zero},
 this sends $p_i(\zeta)$ to 
\begin{equation} B_i \frac{\prod_{j<i}C_{ji}(\zeta-a_{ji})}{\prod_{j>i}C_{ij}(\zeta-a_{ij})} e^{x_i-\bar{z}_i\zeta}=E_i \frac{\prod_{j<i}(\zeta-a_{ji})}{\prod_{j>i}(\zeta-a_{ij})} e^{x_i-\bar{z}_i\zeta},\label{pi}\end{equation}
where
\begin{equation} E_i\bar{E}_i= \frac{\prod_{j<i}(x_i-x_j+r_{ij})}{\prod_{j>i}(x_j-x_i+r_{ij})}. \label{Ei}\end{equation}
These and the $\beta_i$ give the real sections for the Gibbons-Manton metric and the symplectic form is \eqref{om1}. We now compare this with the description of $Z(1)$ given in Proposition \ref{twistor2}. According to Remark \ref{newform} we should set $p_i=u_i\big/\prod_{j>i}(\beta_i-\beta_j)$ in order to have the same symplectic form. We obtain
$$p_i(\zeta)=\frac{A_i}{\prod_{j>i} (\bar{z}_j-\bar{z}_i)} \frac{\prod_{j<i}(\zeta-a_{ji})}{\prod_{j>i}(\zeta-a_{ij})} e^{x_i-\bar{z}_i\zeta}.$$
All we have to do is to compare is the norm of $A_i\big/\prod_{j>i} (\bar{z}_j-\bar{z}_i)$ with the norm of $E_i$. We have, from Proposition \ref{twistor2} and equation \eqref{Ei},
\begin{multline*}\frac{A_i\bar{A}_i}{\prod_{j>i} |z_j-z_i|^2}= \frac{\prod_{j\neq i}(x_i-x_j+r_{ij})}{\prod_{j>i} |z_j-z_i|^2}= \prod_{j< i}(x_i-x_j+r_{ij})\prod_{j> i}\frac{(x_i-x_j+r_{ij})}{|z_j-z_i|^2}\\ = \frac{\prod_{j<i}(x_i-x_j+r_{ij})}{\prod_{j>i}(x_j-x_i+r_{ij})}=E_i\bar{E}_i,
\end{multline*}
which proves the theorem.
\end{pf}

We shall finish the section with a remark that Propositions \ref{twistor2} and \ref{bundle} can be generalized to define hyperk\"ahler metrics on a class of $T^n$-bundles over $\tilde{C}_n({\Bbb R}^3)$. We have:

\begin{theorem} Let $P$ be a $T^n$-bundle over $\tilde{C}_n\bigl({\Bbb R}^3\bigr)$ determined by an element $(s_1,\ldots,s_n)$ of $H^2\bigl(\tilde{C}_n({\Bbb R}^3),{\Bbb Z}^n\bigr)$ satisfying $s_k(S_{ij}^2)=0$ if $k\neq i,j$ and $s_i(S_{ij}^2)=-s_j(S_{ij}^2)$. Then $P$ carries a family of (pseudo)-hyperk\"ahler metrics such that the real sections of the twistor space are given, over $\zeta\neq \infty$, by $(\beta_1(\zeta), \ldots,\beta_n(\zeta),\linebreak u_1(\zeta),\ldots,u_n(\zeta))$, where
\begin{align*} & \beta_i(\zeta)=z_i+2x_i\zeta-\bar{z}_i\zeta^2,\\
&  u_i(\zeta)=A_i\prod_{j\neq i}(\zeta-a_{ji})^{s_{ij}}e^{c(x_i-\bar{z}_i\zeta)},\end{align*}
where $c$ is a real constant, $(x_i,z_i)$, $i=1,\ldots,n$, are distinct points in ${\Bbb R}\times{\Bbb C}$, $s_{ij}=|s_i(S^2_{ij})|$, and $A_i$ are complex numbers satisfying
$$\begin{array}{ccr} \hspace*{3cm} & A_i\bar{A}_i=\prod_{j\neq i}\bigl(x_i-x_j+r_{ij}\bigr)^{s_{ij}}. & \hspace{3.5cm}\Box \end{array}$$
\label{conf}\end{theorem}
This description determines a hypercomplex structure on $P$.
A (pseudo)-hyperk\"ahler metric can be then calculated using any  complex-symplectic form along the fibers, given as a section of $\Lambda^2T_F^\ast\otimes O(2)$, e.g. the form \eqref{omegaf}.
   These metrics will correspond to the motion of $n$ dyons in ${\Bbb R}^3$ interacting in different ways (cf. \cite{GM}).

\begin{remarknon} The calculation of the metric given above shows that the Taub-NUT metric (cf. \cite{Besse}) has two very different descriptions in terms of Nahm's equaations: 1) it is the metric on the totally geodesic submanifold $\tilde{M}_2^0(-1)$ of $\tilde{M}_2(-1)$ defined by considering $\frak{su}(2)$-valued solutions to Nahm's equations and $SU(2)$-valued gauge transformations; 2) it is the metric on the moduli space of $SU(3)$-monopoles of charge $(1,1)$ \cite{Con,Murr}. \end{remarknon}

\section{ Asymptotic comparison of the metrics\label{four}}

We shall now show that the Gibbons-Manton metric and the monopole metric are asymptotically exponentially close. The asymptotic region, where the individual monopoles are separated, of the monopole space $M_n$ is diffeomorphic to $P/S_n$, where $P$ is a torus bundle over the configuration space $\tilde{C}_n({\Bbb R}^3)$ and $S_n$ the symmetric group. The bundle $P$ is not, however, the bundle of Proposition \ref{bundle}. Rather, as we shall see shortly, it is the quotient of that bundle by a $\bigl({\Bbb Z}_2\bigr)^n$-subgroup of $T^n$. In other words it is the bundle determined by an $s\in H^2\bigl(\tilde{C}_n({\Bbb R}^3),{\Bbb Z}^n\bigr)$ with all $s_k$ being twice of those in Proposition \ref{bundle}. 
\par
We shall compare the metric on $M_n$ with the metric on the hyperk\"ahler quotient of $\tilde{M}_n(1) \times \tilde{M}_n(1)$ by the diagonal $T^n$-action. We do this in order to have solutions to Nahm's equations with poles at both ends of the interval $[-1,1]$. For any $c,c^\prime$, let us write $\tilde{M}_n(c,c^\prime)$
for the hyperk\"ahler quotient of $\tilde{M}_n(c)\times \tilde{M}_n(c^\prime)$ by the diagonal action of $T^n$. 
The action of $T^n$ given by $t\cdot(m,m^\prime)=(tm,m^\prime)$ induces a tri-Hamiltonian action of $T^n$ on 
$\tilde{M}_n(c,c^\prime)$ which makes $\tilde{M}_n(c,c^\prime)$ into a $T^n$-bundle over $\tilde{C}_n({\Bbb R}^3)$. 
We have
\begin{lemma}  $\tilde{M}_n(c,c^\prime)$ is isomorphic, as a hyperk\"ahler manifold, to $\tilde{M}_n(c+c^\prime)\big/\bigl({\Bbb Z}_2\bigr)^n$, where $\bigl({\Bbb Z}_2\bigr)^n=\{t\in T^n;t^2=1\}$.\end{lemma}
\begin{pf} Let $\mu,\mu^\prime$ be the moment maps for the action of $T^n$ on $\tilde{M}_n(c),\tilde{M}_n(c^\prime)$ respectively. The moment map for the diagonal $T^n$-action on the product is $\mu+\mu^\prime$. If we go back to the proof of Proposition \ref{bundle} and use the same notation, we can see that the zero-set of this moment map is a $(T^n\times T^n)$-bundle over $\tilde{C}_n({\Bbb R}^3)$ which restricted to each $S^2_{ij}$ is given by transition functions $(\phi_1,\ldots,\phi_n,\phi^{-1}_1,\ldots,\phi^{-1}_n)$ (the point being that $U(I^\prime_1,\ldots,I^\prime_n)=-U(I_1,\ldots,I_n)$). Hence, if we quotient by $T^n$, by sending the second $T^n$ to $1$ over each $U(I_1,\ldots,I_n)$, we end up with a $T^n$-bundle for which the transition functions are $\phi_k^2$, $k=1,\ldots,n$. This proves the differential-geometric part of the statement. To obtain  the isometry we repeat this argument for the twistor space of $\tilde{M}_n(c)\times \tilde{M}_n(c^\prime)$, performing the complex-symplectic quotient along the fibers as in the proof of Theorem \ref{GMmetric}.\end{pf}  

From now on, we shall consider $\tilde{M}_n(1,1)$ with {\em half} (compare the formula \eqref{metric}) of the metric given by the above lemma. In other words, locally the metric is still the Gibbons-Manton metric.
\par
We can identify $\tilde{M}_n(1,1)$ with the moduli space of pairs $\bigl((T_0,T_1,T_2,T_3),\linebreak(T_0^\prime,T_1^\prime,T_2^\prime,T_3^\prime) \bigr)$ of solutions to Nahm's equations, defined respectively on $[-1,\infty]$ and on $[-\infty,1]$, such that $T_i(+\infty)=T_i^\prime(-\infty)$ for $i=0,1,2,3$, and the residues of $T_i$ at $-1$ and of $T^\prime_i$ at $+1$, $i=1,2,3$, define the standard $n$-dimensional irreducible representation of ${\frak su}(2)$. The group of gauge transformations ${\cal G}(1,1)$ is now defined as pairs $(g,g^\prime)$ such that $g(t+1),g^\prime(-t+1)\in {\cal G}(c)$  for some $c$ and $s=\lim_{t\rightarrow +\infty} \dot{g}g^{-1}= \lim_{t\rightarrow -\infty} \dot{g}^\prime g^{\prime-1}$. The tangent space consists of pairs $\bigl((t_0,t_1,t_2,t_3),(t_0^\prime,t_1^\prime,t_2^\prime,t_3^\prime) \bigr)$
defined on $[-1,\infty]$ and on $[-\infty,1]$, respectively, with $t_i(+\infty)=t^\prime_i(-\infty)$ and satisfying equations \eqref{tangent}. The metric on $\tilde{M}_n(1,1)$ can be written as
\begin{multline*}\frac{1}{2}\sum_0^3\|t_i(+\infty)\|^2+\frac{1}{2}\int_{-1}^{+\infty}\sum_0^3\left(\|t_i(s)\|^2-\|t_i(+\infty)\|^2\right)ds+\\  \frac{1}{2}\sum_0^3\|t_i^\prime(-\infty)\|^2+\frac{1}{2}\int_{-\infty}^1\sum_0^3\left(\|t_i^\prime(s)\|^2-\|t_i^\prime(-\infty)\|^2\right)ds. \label{newsmetric}\end{multline*}
 We can rewrite this as
\begin{multline} \frac{1}{2}\int_{0}^{+\infty}\sum_0^3\left(\|t_i(s)\|^2-\|t_i(+\infty)\|^2\right)ds + \frac{1}{2}\int_{-\infty}^0\sum_0^3\left(\|t_i^\prime(s)\|^2-\|t_i(+\infty)\|^2\right)ds\\
+\frac{1}{2}\int_{-1}^{0}\sum_0^3\|t_i(s)\|^2ds+ \frac{1}{2}\int_{0}^1\sum_0^3\|t_i^\prime(s)\|^2ds. \label{supermetric}\end{multline}
Let us fix complex structure, say $I$ and write as in section \ref{twoandhalf}, $\alpha$ for $T_0+iT_1$, $\beta$ for $T_2+iT_3$. We write an element of $\tilde{M}_n(1,1)$ as a pair $\bigl((\alpha_{-},\beta_{-}),(\alpha_{+},\beta_{+})\bigr)$.  We shall write $\beta_i$ for the $(i,i)$-th entry of $\beta_{-}(+\infty)=\beta_{+}(-\infty)$ and denote by $\tilde{M}_n^{\rm reg}(1,1)$ the subset of $\tilde{M}_n(1,1)$ where all $\beta_i$ are distinct. Similarily, we write $M_n^{\rm reg}$ for the subset of $(\alpha,\beta)$ in $(M_n,I)$ where the eigenvalues of $\beta$ are distinct.
We shall prove:
\begin{theorem} There exists a biholomorphism $\phi$ from $\tilde{M}_n^{\rm reg}(1,1)/S_n$ to  $M_n^{\rm reg}$ such that 
\begin{equation}|\phi^\ast g-g^{\prime}|=O(e^{-cR})\label{phi}\end{equation}
where $g,g^{\prime}$ denote the monopole and Gibbons-Manton metric respectively, $c=c(n)$ is a constant, and $R$ is the separation distance of particles in $C_n({\Bbb R}^3)$, i.e. \begin{equation} R=\min \{|{\bf x}_i-{\bf x}_j|; i\neq j\}.\label{R}\end{equation}
 The same estimate holds for the Riemannian curvature tensor.\label{estimates}\end{theorem}

Since such a biholomorphism will be defined for any complex structure and the union of $\tilde{M}_n^{\rm reg}(1,1)$ for different complex structures is all of $\tilde{M}_n(1,1)$, we conclude that the monopole and the Gibbons-Manton metrics are exponentially close in the asymptotic region of the monopole moduli space. 
\par
The remainder of the section is devoted to proving this theorem. We need the following lemma:
\begin{lemma} Let $C>0$. The space $\tilde{M}_n^{\rm reg}(1)$ is biholomorphic to the quotient of the space of solutions $(\alpha,\beta)$ to the equation \eqref{complex} which have the correct boundary behaviour at $t=0$ and are constant (hence diagonal) for $t\geq C$  by the group of complex gauge transformations $g:[0,+\infty)\rightarrow Gl(n,{\Bbb C})$ with $g(0)=1$ and $g(t)=\exp(ht-h)$ for some diagonal $h$ for $t\geq C$.\end{lemma} 
\begin{pf} Let $(\alpha,\beta)$ be an element of $\tilde{M}_n^{\rm reg}(1)$ and let $\alpha_d=\alpha(+\infty)$, $\beta_d=\beta(+\infty)$. According to the proof of Proposition \ref{Mascomplex}, there is unique complex gauge transformation $g$ defined on $[C/2,+\infty)$ with $g(+\infty)=1$ such that $(\alpha,\beta)=g(\alpha_d,\beta_d)$. Let $\hat{g}:[C/2,\infty) \rightarrow Gl(n,{\Bbb C})$ be a smooth path with the values and the first derivatives of $\hat{g}$ and $g$ coinciding at $t=C/2$ and with $\hat{g}(t)=1$ and for $t\geq C$. We obtain a solution $(\hat{\alpha},\hat{\beta})$ to the complex Nahm equation \eqref{complex} by setting
\begin{equation}(\hat{\alpha},\hat{\beta})(t)=\begin{cases} (\alpha,\beta)(t) & \text{if $t<C$}\\ \hat{g}(t)(\alpha_d,\beta_d) & \text{if $t\geq C$}.\end{cases}\label{smoothing}\end{equation}
This is a solution of the type described in the statement of this lemma. The proof of \ref{Mascomplex} shows further that it is only $g(C/2)\exp\{(1-C/2)\alpha_d\}$ (and a solution to \eqref{complex} on $[0,C/2]$) that determines the element of $\tilde{M}_n^{\rm reg}(1)$. Therefore we obtain a well defined holomorphic map from $\tilde{M}_n^{\rm reg}(1)$ to the moduli space described in the statement. Let us define the inverse map.
Let $(\hat{\alpha},\hat{\beta})$ be an element of the moduli space described in the statement. As in \cite{Kron} we can find a bounded complex gauge transformation $g_0$ such that $g_0(\hat{\alpha},\hat{\beta})$ is an element of $\tilde{M}_n^{\rm reg}(1)$.
We can assume that $g_0$ has a limit $h$ at $+\infty$ (this follows from the convexity property of $g_0$ \cite{Don}, since we can assume that $g_0(t)$ is hermitian for all $t$). According to Proposition \ref{twistor2} the action of $T^n$ on $\tilde{M}_n(1)$ extends to a global action $\bigl({\Bbb C}^\ast\bigr)^n$ with respect to the complex structure $I$ (or any other). Let $(\alpha,\beta)$ be the element of $\tilde{M}_n^{\rm reg}(1)$ obtained from $g_0(\hat{\alpha},\hat{\beta})$ by the action of $h^{-1} \in\bigl({\Bbb C}^\ast\bigr)^n$. Then $(\alpha,\beta)= g(\hat{\alpha},\hat{\beta})$ and $g\in{\cal G}^{\Bbb C}(1)$. This gives the inverse mapping.
\end{pf}

We can now construct a biholomorphism between  $\tilde{M}_n^{\rm reg}(1,1)/S_n$ and $M_n^{\rm reg}$. From the above lemma,  $\tilde{M}_n^{\rm reg}(1,1)$ is biholomorphic to the quotient of the space of pairs $\bigl((\alpha_{-},\beta_{-}),(\alpha_{+},\beta_{+})\bigr)$ defined on $[-1,+\infty)$ and on $(-\infty,1]$ respectively such that $(\alpha_{-},\beta_{-})(t+1)$ and $(\alpha_{+},\beta_{+})(1-t)$ are as in the above lemma, $(\alpha_{-},\beta_{-})(+\infty)=(\alpha_{+},\beta_{+})(-\infty)$ by the group of pairs $(g_{-},g_{+})$ with $g_-(-1)=g_+(1)=1$ and such that there are diagonal $h,p$ with $g_-(t)=\exp(th-p)$ for $t>-r$, $g_+(t)=\exp(th-p)$ for $t<r$ ($r\in(0,1)$ is fixed but arbitrary). We define a solution $(\hat{\alpha},\hat{\beta})$ to the complex Nahm equation \eqref{complex} on $(-1,1)$ by
\begin{equation} (\hat{\alpha},\hat{\beta})(t)=\begin{cases} (\alpha_{-},\beta_{-})(t) & \text{if $t<0$}\\ (\alpha_{+},\beta_{+})(t) & \text{if $t\geq 0$}.\end{cases} \label{hats}\end{equation}
The ${\cal G}^{\Bbb C}$-orbit of this solution (see section \ref{one} for the definition of ${\cal G}$) contains a unique element of $M_n$ \cite{Don,Hur}. Furthermore, the action of a $(g_{-},g_{+})$ translates into the action of $g\in {\cal G}^{\Bbb C}$, where $g(t)=g_-(t)$ for $t<0$ and $g(t)=g_+(t)$ for $t\geq 0$. Therefore we have a well defined holomorphic map $\phi_r$ from $\tilde{M}_n^{\rm reg}(1,1)$ to $M_n$. If we now have an element $(\alpha,\beta)$ of $M_n^{\rm reg}$, we can diagonalize $\beta$ on $[-r,r]$ and make $\alpha$ diagonal and constant on $[-r,r]$. Let $(\tilde{\alpha},\tilde{\beta})$ be the resulting solution to the complex Nahm equation. We obtain an element of $\tilde{M}_n^{\rm reg}(1,1)$ by setting $$(\alpha_{-},\beta_{-})(t)=\begin{cases}(\tilde{\alpha},\tilde{\beta})(t) & \text{for $t<0$}\\ (\tilde{\alpha},\tilde{\beta})(0) & \text{for $t\geq 0$}\end{cases}$$
and similarily for $(\alpha_{+},\beta_{+})$. This defines the inverse to $\phi_r$ up to the ordering of eigenvalues of $\beta$. In other words $\phi_r$ induces a biholomorphism between  $\tilde{M}_n^{\rm reg}(1,1)/S_n$ and $M_n^{\rm reg}$. Furthermore, for a fixed element $\bigl((\alpha_{-},\beta_{-}),(\alpha_{+},\beta_{+})\bigr)$ of  $\tilde{M}_n^{\rm reg}(1,1)$ and two parameters $r,r^\prime$, the resulting $(\hat{\alpha},\hat{\beta})$ of \eqref{hats} are ${\cal G}^{\Bbb C}$-equivalent and therefore $\phi_r,\phi_{r^\prime}$ induce the same biholomorphism $\phi$.
\bigskip

Let us now prove the estimate \eqref{R}. Fortunately, much of the analysis has been already done in \cite{BielCrelle}. First of all, we recall (\cite{Kron}, Lemma 3.4) that solutions to Nahm's equations which have a regular triple as a limit at infinity, approach this limit exponentially fast, of order $O\left(e^{-cR}\right)$ (that is $T_1,T_2,T_3$ do and we can always make $T_0$ to have such decay by using the gauge freedom). The proofs of Propositions 3.11 -- 3.14 in \cite{BielCrelle} show that the same holds for tangent vectors $(t_0,t_1,t_2,t_3)$. Let us now see what happens to a tangent vector $v$ under the map $\phi$. The gauge transformations $(g,g^\prime)$ which make the element $\bigl((\alpha_{-},\beta_{-}),(\alpha_{+},\beta_{+})\bigr)$ of $\tilde{M}_n^{\rm reg}(1,1)$ constant and equal to the common value at infinity on $[-1+C/2,+\infty)$ and $(-\infty,1-C/2]$ are exponentially close to the identity. In the next stage of the construction of $\phi$ - the formula \eqref{smoothing} - we have smoothed out the solutions which can be again done by gauge transformations exponentially close to $1$. Therefore the resulting tangent vector $\hat{v}$ is exponentially close to the original one in the metric \eqref{supermetric}. We have then restricted the solutions (formula \eqref{hats}) to obtain a solution $(\hat{\alpha},\hat{\beta})$ to the complex Nahm equation on $[-1,1]$. Let $p$ denote this  operation of restriction. 
The first line of the formula \eqref{supermetric} is exponentially small and therefore the norm of $\hat{v}$ in \eqref{supermetric} and the norm of $dp(\hat{v})$ in \eqref{metric} are exponentially close. 
 The solution $(\hat{\alpha},\hat{\beta})$ will not satisfy the real Nahm equation, however, we will have
$$F(\hat{\alpha},\hat{\beta}):=\frac{d\,}{dt}\left(\hat{\alpha}+ \hat{\alpha}^\ast\right)+[\hat{\alpha},\hat{\alpha}^\ast]+ [\hat{\beta},\hat{\beta}^\ast]=O(e^{-cR}).$$
Lemma 2.10 in \cite{Don} implies now that we can solve the real equation by a complex gauge transformation bounded as $O(e^{-cR})$. We can now show that the vector $d\phi(v)$ tangent to $M_n$ (which is obtained from $dp(\hat{v})$) is exponentially close to $dp(\hat{v})$ by following the analysis of section 3 in \cite{BielCrelle} step by step, replacing the $O(1/R)$ estimates by $O(e^{-cR})$. This proves the estimate \eqref{R}. For the curvature estimates we do the same using the analysis of section 4 in \cite{BielCrelle}. This proves Theorem \ref{estimates}.

\section{Twistor description of monopoles and the Gibbons-Manton metric}

We shall show in this section how the twistor description of monopole metrics determines the asymptotic metric. We recall \cite{Don} that the moduli space of $n$-monopoles is biholomorphic to the space of based rational maps $p(z)/q(z)$ on ${\Bbb C}P^1$ of degree $n$ (based means that $\text{deg}\,p<\text{deg}\,q$).
On the set, where the roots $\beta_1,\ldots,\beta_n$ of $q(z)$ are distinct, these roots and the values $p_i=p(\beta_i)$ of $p$ form local coordinates and the complex-symplectic form can be written as \cite{AtHi}:
\begin{equation}\sum_{i=1}^n  d\beta_i \wedge\frac{dp_i}{p_i}. \label{formulka} \end{equation}
The metric is determined by the real sections $p(z,\zeta)/q(z,\zeta)$. Their description is provided in \cite{Hurt}. The denominator $q(z,\zeta)$ is given by a curve $S$ - the spectral curve of the monopole  - in $T{\Bbb C}P^1$ \cite{Hit}. This curve satisfies several conditions one of which is the triviality of the line bundle $L^{-2}$ restricted to $S$ and Hurtubise \cite{Hurt} shows that the numerator $p(z,\zeta)$ is given by a nonzero section of this bundle (the values $p_i(\zeta)$ are given by the values of this section at the intersection points $\beta_i(\zeta)$ of $S$ with $T_\zeta{\Bbb C}P^1$.
\par
What happens when the individual monopoles separate? First of all, the spectral curve approaches the union of spectral curves of individual monopoles exponentially fast \cite{BielAGAG}. These curves $S_i$  are of the form $\eta_i=z_i+2x_i\zeta-\bar{z}_i\zeta^2$, $i=1,\ldots,n$, where $(x_i,\text{Re}\,z_i,\text{Im}\,z_i)$ are locations of $1$-monopoles (particles).  What happens to the section of $L^{-2}$? We make a heuristic assumption (which we know to be true from section \ref{three}) that the section acquires zeros and poles at the intersection points of the $S_i$ (more precisely the only singularities of $p_i(\zeta)$ occur at the intersection points of $S_i$ with other $S_j$). As we shall see this is sufficient to determine the asymptotic metric.
\par
First of all the real structure on the bundle $L^{-2}$ is $u\mapsto \bar{u}^{-1}e^{-2\bar{\eta}/\bar{\zeta}}$ and therefore if $p_i$ has a zero at one of the points of $S_i\cap S_j$, then it has a pole of the same order at the other, and vice versa. Furthermore, since the metric and hence the real sections are invariant under the action of the symmetric group, we must have
$$p_i(\zeta)=A_i\prod_{j\neq i}\left(\frac{\zeta-a_{ij}}{\zeta-a_{ji}}\right)^k  e^{-2(x_i-\bar{z}_i\zeta)},\qquad i=1,\ldots,n,$$
where $a_{ij},a_{ji}$ are the two points in $S_i\cap S_j$ given by \eqref{aij} and $k$ is an integer. The reality condition implies that
$$ A_i\bar{A}_i= \prod_{j\neq i}a_{ji}^k\bar{a}_{ji}^k.$$
One can now calculate the asymptotic metric, using \eqref{formulka}. The sign of $k$ will determine the signature, while $|k|$ is simply a constant multiple. The actual value of $k$ is determined by the topology of the asymptotic region of $M_n$, and comparing with Proposition \ref{bundle} and the remarks at the beginning of section \ref{four} we conclude that $k=1$ (in the coordinates of Proposition \ref{twistor2}, $p_i=\prod_{j;j\neq i}(\beta_i-\beta_j)/u^2_i$).
\bigskip

We remark that the above analysis can be easily done for other compact Lie groups $G$. The twistor description of metrics on moduli spaces of $G$-monopoles with maximal symmetry breaking is known from the work of Murray \cite{Mur} and Hurtubise and Murray \cite{HurtMur, HurMur} and from this the asymptotic metric can be calculated. We shall do the exact analysis in the case of $G=SU(N)$ in a subsequent paper.


\begin{thebibliography}{99}


 
\bibitem{AtHi}
{ M.F. Atiyah  \and N.J. Hitchin}, {\em The geometry and dynamics of magnetic monopoles}, Princeton University Press,  Princeton (1988).

\bibitem{Besse}
{ A. Besse}, {\em Einstein manifolds}, Springer (1987).

\bibitem{BielCrelle}
{ R. Bielawski}, `Asymptotic behaviour of $SU(2)$ monopole metrics', {\it J. reine angew. Math.}, 468 (1995), 139--165.

\bibitem{BielAGAG}
{ R. Bielawski}, `Monopoles, particles and rational functions', {\it Ann. Glob. Anal. Geom.}  14 (1996), 123--145.

\bibitem{BielAGAG2}
{ R. Bielawski}, `On the hyperk\"ahler metrics associated to singularities of nilpotent varieties', {\it Ann. Glob. Anal. Geom.}  14 (1996), 177--191.

\bibitem{BielJLMS}
{ R. Bielawski}, `Hyperk\"ahler structures and group actions', {\it J. London Math. Soc.}, 55 (1997), 400--414.

\bibitem{BielCam}
R. Bielawski, `Invariant hyperk\"ahler metrics with a homogeneous complex structure', {\it Math. Proc. Cam. Phil. Soc.}, to appear.

\bibitem{Biq}
{ O. Biquard}, `Sur les \'equations de Nahm et les orbites coadjointes des groupes de Lie semi-simples complexes', {\it Math. Ann.} 304 (1996), 253--276.

\bibitem{Biq2}
{ O. Biquard}, `Twisteurs des orbites coadjointes', Ecole Polytechnique preprint (1997).

\bibitem{Con}
S. Connell, `The dynamics of the $SU(3)$ $(1,1)$ magnetic monopoles', Ph.D. thesis, The Flinders University of South Australia (1991).

\bibitem{Dan1}
{ A.S. Dancer}, `Nahm's equations and hyperk\"ahler geometry', {\it Commun. Math. Phys.} 158 (1993), 545--568.

\bibitem{Dan2}
{ A.S. Dancer}, `A family of hyperk\"ahler manifolds', {\it Quart. J. Math. Oxford} 45 (1994), 463--478.

\bibitem{Don}
{ S.K. Donaldson}, `Nahm's equations and the classification of monopoles', {\it Commun. Math. Phys.} 96 (1984), 387--407.

\bibitem{GM}
{ G.W. Gibbons \and N.S. Manton}, `The moduli space metric for well-separated BPS monopoles', {\it Phys. Lett. B} 356 (1995), 32--38.

\bibitem{GR}
{ G.W. Gibbons \and P. Rychenkova}, `HyperK\"ahler quotient construction of BPS monopole moduli spaces', {\it Commun. Math. Phys.} 186 (1997), 581--599.

\bibitem{Hit}
{ N.J. Hitchin}, `On the construction of
monopoles',  {\it Commun. Math. Phys.} 89 (1983), 145--190.

\bibitem{HitCam}
N.J. Hitchin, `Polygons and gravitons', {\it Math. Proc. Camb. Phil. Soc.} 83 (1979), 465--476.

\bibitem{HKLR}
{ N.J. Hitchin, A. Karlhede, U. Lindstr\"om, \and M. Ro\v{c}ek}, `Hyperk\"{a}hler metrics and supersymmetry',  {\it Commun. Math. Phys.} 108 (1985), 535--586.

\bibitem{Hurt}
{ J.C. Hurtubise}, `Monopoles and rational maps: a note on a theorem of Donaldson', {\it Commun. Math. Phys.} 100 (1985), 191--196.

\bibitem{Hur}
{ J.C. Hurtubise}, `The classification of monopoles for the classical groups', {\it Commun. Math. Phys.} 120 (1989), 613--641.

\bibitem{HurtMur}
{ J.C. Hurtubise \and M.K. Murray}, `On the construction of monopoles for the classical groups',  {\it Commun. Math. Phys.} 122 (1989), 35--89.

\bibitem{HurMur}
{ J.C. Hurtubise \and M.K. Murray}, `Monopoles and their spectral data' 
{\it Commun. Math. Phys.} 133 (1990), 487--508.

\bibitem{Kron}
{ P.B. Kronheimer}, `A hyper-k\"ahlerian structure on coadjoint orbits of a semisimple complex group', {\it J. London Math. Soc.} 42 (1990), 193--208.

\bibitem{KronALE}
{ P.B. Kronheimer}, `The construction of ALE spaces as hyper-K\"ahler quotients', {\it J. Diff. Geom.} 29 (1989), 665--683.

\bibitem{LR} 
{ U. Lindstr\"om
\and M. Ro\v{c}ek}, `Scalar tensor duality and $N=1,2$ nonlinear
$\sigma$-models', {\it Nucl. Phys.} 222B (1983), 285-308. 

\bibitem{Man}
{ N.S. Manton}, `A remark on the scattering of BPS monopoles', {\it Phys. Lett. B} 110 (1982), 54--56.

\bibitem{Man1}
{ N.S. Manton}, `Monopole interactions at long range', {\it Phys. Lett. B} 154 (1985), 397--400.

\bibitem{Mur}
{ M.K. Murray}, `Non-abelian magnetic monopoles', {\it 
Commun. Math. Phys.} 96 (1984), 539--565.

\bibitem{Murr}
{ M.K. Murray}, `A note on the $(1,1,\ldots,1)$ monopole metric', {\it J. Geom. Phys.} 23 (1997), 31--41.

\bibitem{Nahm}
{ W. Nahm}, `The construction of all self-dual monopoles by the ADHM method', in {\em Monopoles in quantum field theory}, World Scientific, Singapore (1982).


\bibitem{Nak}
{ H. Nakajima},  `Monopoles and Nahm's equations' in {\em Einstein metrics and Yang-Mills connections}, Marcel Dekker, New York (1993).

\bibitem{PP}
{ H. Pedersen \and Y.S. Poon}, `Hyper-K\"ahler metrics and a
generalization of the Bogomolny equations,' {\it Comm. Math. Phys.} 
117 (1988), 569--580. 

\bibitem{Stu}
{ D. Stuart}, `The geodesic approximation for the Yang-Mills-Higgs equations', {\it  Commun. Math. Phys.} 166 (1994), 149--190.

\bibitem{Swann}
{ A. Swann}, `Hyperk\"ahler and quaternionic K\"ahler geometry', {\it Math. Ann.} 289 (1991), 421--450.

\end{thebibliography}
\end{document}